\documentclass[pdflatex,sn-mathphys-ay]{sn-jnl}

\usepackage[T1]{fontenc}
\usepackage{graphicx}
\usepackage{amsmath,amssymb}
\usepackage{booktabs}
\usepackage{subcaption}
\usepackage{enumitem}
\usepackage[table]{xcolor}
\usepackage{tabularx}
\usepackage{array}
\usepackage{makecell}
\usepackage{pdflscape}
\usepackage[title]{appendix}
\usepackage{hyperref}

\graphicspath{{./}{figures/}}

\begin{document}
\title{A Programming, Robotics and Automatic Control Intervention for Secondary Education Students in Sweden and Italy}

\author*[1]{\fnm{Andre Shigueo} \sur{Yamashita-Fridesjö}}\email{andre.yamashita@optimation.se}
\author[2]{\fnm{Carolina} \sur{Hertzberg}}\email{carolina.hertzberg@optimation.se}

\author[3]{\fnm{Martina} \sur{Morano}}\email{m.morano@pm.univpm.it}

\author[4]{\fnm{Giacomo} \sur{Fiara}}\email{g.fiara@ancybernetics.it}

\author[4]{\fnm{Benedetta} \sur{Castagna}}\email{b.castagna@ancybernetics.it}

\author[3]{\fnm{David} \sur{Scaradozzi}}\email{d.scaradozzi@staff.univpm.it}



\affil*[1]{\orgdiv{Decision support}, \orgname{Optimation AB}, \orgaddress{\street{Varvsgatan 53}, \city{Luleå}, \postcode{97233}, \country{Sweden}}}

\affil[2]{\orgdiv{Advanced process control}, \orgname{Optimation AB}, \orgaddress{\street{Eklundshovsvägen 1H}, \city{Uppsala}, \postcode{75237}, \country{Sweden}}}

\affil[3]{\orgdiv{Department of Information Engineering}, \orgname{Università Politecnica delle Marche}, \orgaddress{\city{Ancona}, \country{Italy}}}

\affil[4]{\orgname{ANcybernetics}, \orgaddress{\city{Ancona}, \country{Italy}}}



\abstract{This study evaluates the AutoSTEM workshop, an  intervention for secondary education that combines a serious game with PID-controlled hardware. Seven workshops were conducted across Sweden and Italy. The six Swedish workshops adopted an explicitly SDT-informed facilitation framework, while the Italian workshop followed the existing educational robotics format without a formally defined SDT facilitation protocol. Self-reports, an objective knowledge quiz, and qualitative responses were analysed using multiple site-adjusted statistical approaches. Technical-specific self-efficacy increased substantially in both conditions, with large effect sizes, while objective knowledge showed only modest, non-significant gains. No significant site-adjusted differences were found between SDT and control conditions for the primary outcomes. Perceived engagement was higher in the control condition, although this finding should be interpreted cautiously. The objective knowledge composite remained positive but non-significant across sensitivity analyses. Female participants reported smaller PID self-efficacy gains than males. Qualitative findings supported evidence of PID learning while identifying technical and pacing barriers. Overall, the results suggest that the technology-rich intervention supports competence development, regardless of SDT-informed facilitation.
}

\maketitle

\section{Introduction}
\label{sec:introduction}

The Fourth Industrial Revolution has fundamentally restructured the global industrial
landscape through the integration of cyber-physical systems, artificial intelligence, and
robotics. This transformation has generated an acute skills gap, with
approximately 87\% of manufacturing companies worldwide reporting severe difficulties
recruiting personnel with competencies in industrial automation
and robotics systems integration~\citep{urrea2025industrial}. The educational robotics
market alone is projected to grow from \$1.9~billion in 2025 to \$7.9~billion by
2035, underscoring the urgency of developing a workforce capable of meeting the demands
of a new global economy~\citep{chakraborty2024industry}. Furthermore, despite comprising over half of the undergraduate population in OECD (Organisation for Economic Co-operation and Development) countries, women account for only 16\% of STEM university enrolments in the European Union, and the age range of 14 to 18 represents a critical window for intervention, as career attitudes become prominent during secondary education~\citep{lopez2023understanding}.

The AutoSTEM project was conceived within this dual context---the need to increase automation awareness
and the imperative to close the STEM gender gap. It is an Erasmus+ KA210-SCH small-scale
partnership bringing together a transnational consortium with partners from Italy and Sweden. The intervention consists of a two-step simulation-to-real framework combining a simulator with a self-balancing robotics platform, designed to make PID control accessible to secondary school students aged 14--18. The serious game at the core of the simulation step, BRIDGE, and its preliminary validation have been described\citep{morano2026serious}. The present study evaluates the complete intervention across the workshop sessions considered here, with particular focus on how aspects related to participant engagement and motivation may be enhanced using Self-Determination Theory, a psychological framework used to understand human motivation and personality. . The main results are as follows: (i) the AutoSTEM evaluation is built on three complementary evidence sources: self-report Likert items, an {objectively scored} multiple-choice
knowledge quiz, and thematically coded open-ended responses; (ii) it applies a
layered site-adjustment strategy---session fixed-effects models, multilevel models with
session random intercepts, covariate-adjusted difference-in-differences estimation, and
within-country moderator analyses---to separate condition-related change from stable
site characteristics, and it reports leave-one-session-out sensitivity analyses that
quantify the robustness of the pooled estimates to individual sites; (iii) it provides
an honest assessment of a theory-informed instructional contrast (Self Determination Theory-guided
facilitation in Sweden versus delivery in Italy without a formally defined SDT protocol) under the recognition that
this contrast remains confounded by site, and it derives design recommendations for
future multi-site educational technology evaluations.

\subsection{Research Questions}
\label{sec:rqs}

The study is guided by four research questions:

\begin{enumerate}
  \item Do students show pre--post gains in PID-specific self-efficacy, control-system
  understanding, and STEM interest following the AutoSTEM workshop, and are these gains
  robust across sites?
  \item Do gains on self-reported competence converge with gains on an objectively
  scored knowledge quiz and with qualitative accounts of learning?
  \item After controlling for site-specific differences, does SDT-informed facilitation
  produce different outcomes compared to a comparison condition without an explicit SDT-informed facilitation protocol of the same technical content?
  \item Do outcomes differ by gender, and are gendered patterns consistent across
  conditions?
\end{enumerate}

And aims to fill the following research gaps identified in the literature:
\begin{enumerate}
    \item \textbf{Multi-method evidence in educational robotics.} This study combines self-report,
    objectively scored knowledge, and qualitative evidence, and reports diagnostics of
    common-method variance to evaluate an intervention with a small sample size and substantial confounded variables.

    \item \textbf{Site-adjusted estimation in multi-site interventions.} This study applies session fixed-effects, multilevel,
    covariate-adjusted, and within-country sensitivity analyses and reports all
    between-condition contrasts under each specification.

    \item \textbf{SDT research in educational robotics and control engineering.} The
    application of SDT to secondary-school control engineering remains largely
    unexplored; the present study contributes evidence on SDT-informed facilitation in
    this domain while explicitly acknowledging the limitations of its quasi-experimental
    contrast.

    \item \textbf{Gender-related effects.} The interaction between instructional
    approach, facilitation, and gender-specific outcomes such as self-efficacy remains
    underexplored; gender effects are examined here with site-adjusted models.

    \item \textbf{Empirical evidence on serious games.} The B.R.I.D.G.E. (Boosting Robotics Interaction and Dynamic Guidance through Entertainment) simulator provides an opportunity to examine whether educational technology can support competence development even when facilitator-level SDT strategies differ.
\end{enumerate}

The remainder of this paper is structured as follows. Section~\ref{sec:litreview} synthesises the relevant literature. Section~\ref{sec:methodology} describes the intervention architecture, the study design, the instruments, and the site-adjusted analysis strategy. Section~\ref{sec:results} reports the findings. Section~\ref{sec:discussion} interprets the results and discusses implications for evaluation practice, and Section~\ref{sec:conclusion} concludes the paper.

\section{Literature Review}
\label{sec:litreview}

\subsection{Background}
\label{sec:lit-background}

The transition from Industry~4.0 to Industry~5.0 has increased demand for engineers who possess both advanced technical competencies and human-centred skills such as collaboration, adaptability, and problem solving \citep{chakraborty2024industry}. Research highlights a persistent gap between educational robotics and industrial practice, with most educational activities relying on simplified robotics platforms despite evidence that exposure to more advanced technologies improves transferable skills and workforce readiness \citep{urrea2025industrial}. Effective engineering education therefore increasingly emphasises authentic learning environments, active participation, real-world applications, and iterative design processes \citep{ramirez2024evaluation}. Control-engineering interventions provide a particularly promising vehicle because it combines mathematical reasoning, programming, and tangible experimentation~\citep{sapi2025control}.

A substantial body of research demonstrates that STEM interventions can positively influence students' attitudes, self-efficacy, and career interests. Inquiry-based and project-based approaches, particularly when combined with robotics and hands-on activities, have been shown to increase engagement and strengthen interest in STEM careers \citep{ribeirinha2024investigating}. Out-of-school STEM programmes further benefit participants through mentorship, role models, and culturally relevant learning experiences, with particularly strong effects for underrepresented groups \citep{dresser2024increasing, browning2025three}. Across studies, effective interventions consistently combine active learning, psychosocial support, mentorship opportunities, and rigorous evaluation methods \citep{kier2014development}. Mixed-method designs that combine quantitative outcome measures with qualitative process evidence are increasingly recommended in this literature \citep{fabregues2023use}.

Gender disparities remain a major challenge in STEM education and career participation \citep{stoet2018gender, parmaxi2023gender}. Social Cognitive Career Theory (SCCT) suggests that contextual support, role models, and self-efficacy play especially important roles in shaping female students' STEM aspirations \citep{wang2023gender}. Research highlights the importance of addressing barriers such as low sense of belonging and impostor feelings while promoting equitable participation in technical activities \citep{hernandez2024challenge, ditullio2019gender}. Consequently, evidence-based strategies include early interventions, exposure to female role models, inclusive learning environments, and pedagogical approaches that actively support the participation and confidence of underrepresented groups \citep{kovaleva2024pros}.

\subsection{Self-Determination Theory in STEM Education}
\label{sec:lit-sdt}

Self Determination Theory (SDT) explains motivation through the satisfaction of three basic psychological needs: autonomy, competence, and relatedness \citep{ryan2000self, deci2008self}. When these needs are supported, learners develop autonomous forms of motivation associated with greater persistence, creativity, engagement, and well-being. Conversely, need frustration can lead to controlled motivation, disengagement, and poorer learning outcomes \citep{ryan2020intrinsic}. As motivation often declines during adolescence, SDT has become an increasingly important framework for understanding and improving learning experiences in secondary STEM education \citep{gilroy2026self}.

Research in secondary school settings shows that effective SDT-based teaching requires a balance between autonomy support and structure, allowing students to develop competence while gradually increasing their independence \citep{cheon2020autonomy, gilroy2026self}. Empirical studies demonstrate that autonomy-supportive teaching enhances autonomous motivation, effort regulation, and academic achievement, while a large meta-analysis confirmed that autonomous motivation is consistently associated with positive educational outcomes \citep{leon2015self, howard2021student}. Longitudinal evidence further suggests that autonomy support can compensate for lower levels of creativity, confidence, or prior preparation, helping diverse groups of students remain engaged and successful in STEM learning environments \citep{nunez2025how, wang_autonomy-supportive_2025}. The relevance of SDT extends to technology-enhanced learning, vocational education, and educational robotics. Studies indicate that serious games and hands-on learning environments designed to support autonomy, competence, and relatedness can improve both motivation and learning outcomes \citep{farrell_applying_2014}. Similarly, facilitator behaviour, mentoring, and instructional design play critical roles in shaping learner engagement \citep{suyatmo2025preparing}. Within educational robotics, greater student autonomy is generally associated with improved achievement, confidence, and enjoyment, although some learners may initially require additional structure before progressing toward more independent learning \citep{andic2024direct}. Together, these findings support SDT-informed approaches that combine structured guidance, progressive autonomy, and attention to individual learner needs in STEM education \citep{alston2026business}.

\subsubsection{SDT and SCCT: A Complementary Framework for STEM Intervention Design}
\label{sec:lit-scct}

Social Cognitive Career Theory (SCCT) and SDT offer complementary perspectives on STEM career development that, when
integrated, enable a more complete account of educational intervention
effects~\citep{kier2014development, leon2015self}. SCCT identifies the constructs that interventions should target: self-efficacy, outcome expectations, personal goals, and contextual supports. SDT specifies the instructional conditions under which these constructs develop and are internalised: self-efficacy beliefs are strengthened in environments that provide optimal challenge and competence-affirming feedback (competence need satisfaction); outcome expectations are internalised when students experience content as personally meaningful and volitionally chosen (autonomy need satisfaction); and contextual supports exert their influence through the satisfaction of belonging needs (relatedness need satisfaction)~\citep{niemiec2009autonomy, vansteenkiste2020basic}. The results presented here use questionnaire measures interpreted through an SCCT framework to measure intervention outcomes and SDT principles to design the instructional conditions hypothesised to produce them.
The contrast between SDT-supported and comparison condition workshops provides an
exploratory test of whether autonomy-supportive instructional design---over and above
the technical content itself---produces differential effects on the SCCT constructs
that underpin STEM career development.

\section{Methodology}
\label{sec:methodology}

\subsection{The AutoSTEM Intervention}
\label{sec:meth-intervention}

The AutoSTEM intervention is a multi-component programme designed to introduce
secondary school students aged 14--18 to automatic control and robotics. The
workshop duration is approximately three to six hours, depending on the availability of the school, structured around three
topics: (1) Fundamentals of Automatic Control Systems, Robotics and Programming; (2) Learning PID with BRIDGE; and (3) Hands-On Laboratory. 

\subsubsection{Fundamentals of Automatic Control Systems, Robotics and Programming}

The first part of the workshop contains the theory and concepts that are important and relevant for students with little or no knowledge about robotics, programming and automatic control. Figure \ref{fig:material1_agreements} shows how the workshop rules presented in the Swedish workshops within an SDT-informed framework, highlighting the students’ autonomy, relatedness and competence. Some parts of the material are complex for secondary education students, requiring higher levels of abstraction, generalisation, and interpretation of system behaviour. The following concepts in automatic control were introduced, in the following order:
\begin{itemize}
  \item process
  \item input (manipulated variable) and output (measured variable)
  \item block diagram and black-box model
  \item process types (stable, unstable, oscillatory)
  \item dynamic and static responses, steady-state and transient responses
  \item system as a combination of processes
  \item error and setpoint
  \item sensor, controller, actuator
  \item open-loop control and disturbance
  \item controller performance and overshoot
  \item closed-loop control
  \item PID controller and its tuning parameters
\end{itemize}

\begin{figure}[htbp]
\centering
\includegraphics[width=\textwidth]{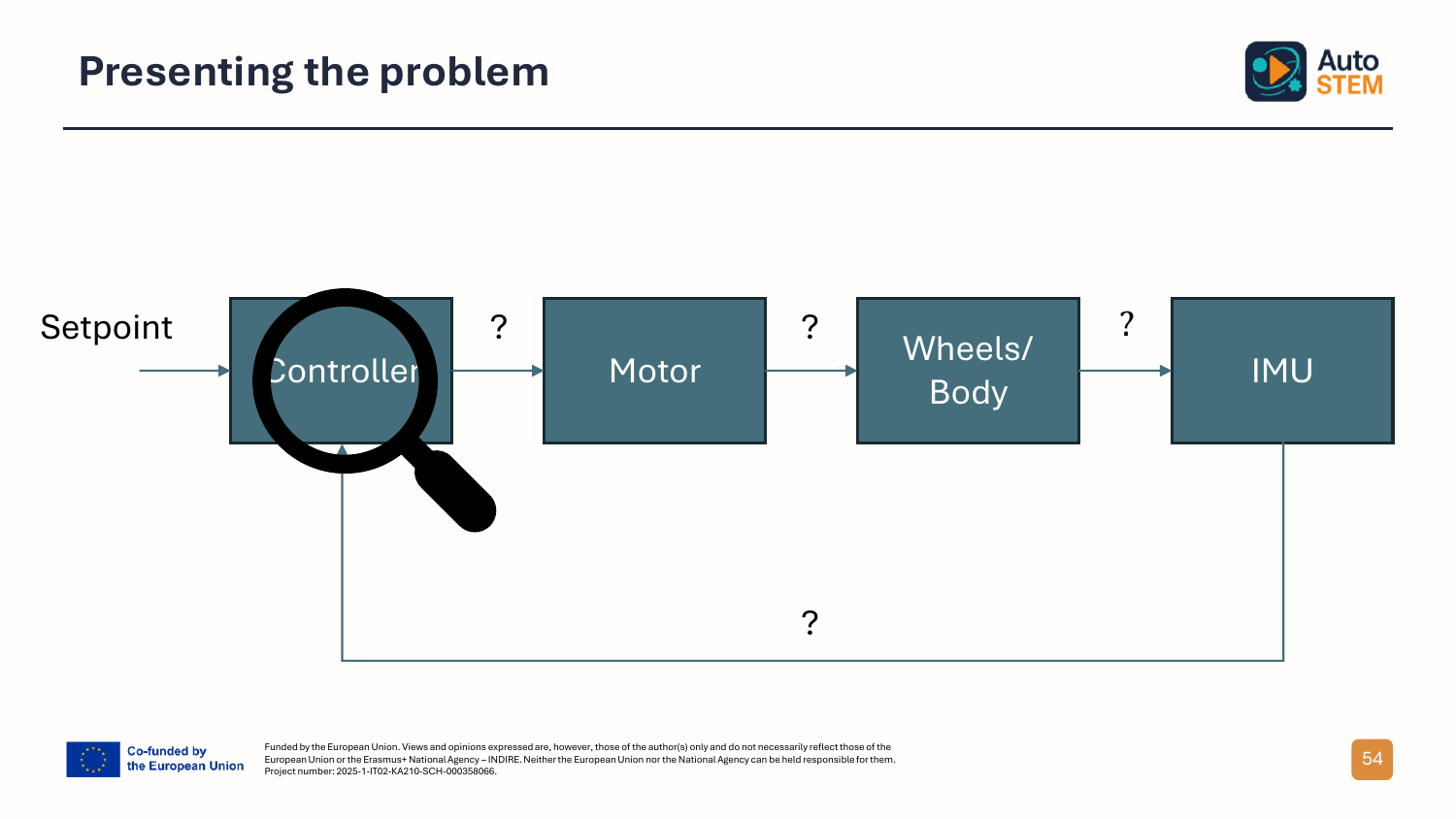}
\caption{Block diagram analysis of the balancing robot problem.}
\label{fig:material1_blockdiagram}
\end{figure}

Figure \ref{fig:material1_blockdiagram} illustrates the block diagram of the control problem of balancing an unstable robot. Through the lecture, other applications of control systems, with industrial, academic, entertainment and personal relevance are presented.
\begin{figure}[ht]
\centering
\fbox{
\parbox{0.9\linewidth}{
\textbf{Our agreements}

\begin{itemize}
  \item It's My Choice: I can choose how to join the activity. It is always okay to take a quiet break and just watch if I need to.
  \item Ask Why: If I don't understand why we are doing something, I can always ask 'why?'
  \item Great Ideas Everywhere: We take turns listening, because everyone here has great ideas and we want to hear them all.
  \item Oops!: Mistakes are just how our brains learn! When things go wrong, we work together to try again instead of getting upset.
  \item We Care: The material we are using was used and will be used by other students. We take care of it. We use the robots for the exercises, not as toys.
\end{itemize}
}}
\caption{Excerpt from Fundamentals of Robotics, Programming and Automatic Control Systems presenting what to expect during the workshop.}
\label{fig:material1_agreements}
\end{figure}

\begin{figure}[htbp]
\centering
\includegraphics[width=\textwidth]{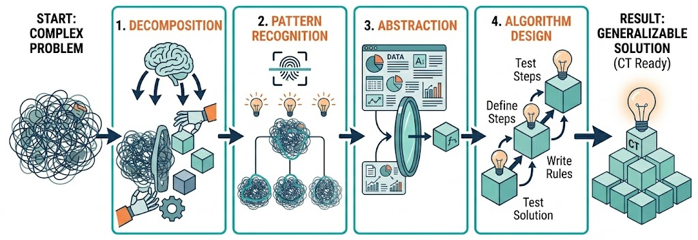}
\caption{Presentation of the Computational Thinking method.}
\label{fig:material1_computationalthinking}
\end{figure}

Besides concepts in automatic control, the workshop relies on a basic knowledge about programming and problem solving, which was presented as Computational Thinking, a methodology used to break down large problems into more tractable parts, and step-by-step solve the smaller problems in a structured way, illustrated in Figure \ref{fig:material1_computationalthinking}. Finally, a more detailed explanation about the sensors, actuators and controller implemented in the robotic kit used later in the workshop are presented, according to Figure \ref{fig:material1_robotikkit}.

\begin{figure}[ht]
\centering
\fbox{
\parbox{0.9\linewidth}{
\textbf{Microcontrollers, sensors, actuators}
\begin{itemize}
  \item Sensors = input (measure the world) temperature, light, distance, sound, motion (IMU)
  \item Microcontroller = the ``brain'' (runs code in a loop) reads sensors $\rightarrow$ makes a decision $\rightarrow$ sends commands
  \item Actuators = output (change the world) motors/servos, LEDs, buzzer/speaker, heater, valve
\end{itemize}

\vspace{1cm}
\textit{Sensors give information from the real world. The microcontroller reads that data and runs the program in a repeating loop. Actuators are the parts that do something physical---move, light up, make sound, heat, or open/close.}

}}
\caption{Sensors, actuators and controller in the context of the robotic kit.}
\label{fig:material1_robotikkit}
\end{figure}

\subsubsection{Learning PID with BRIDGE}
The goals of this part are to explain the advantages and disadvantages of simulators, and how professionals use simulators, before starting the activities with BRIDGE, and then to introduce the PID controller and its parameters. The concept of dynamic simulators is presented, as well as its uses in the process industry: identify bottlenecks, estimate consumption of wood, water and electricity and production of paper, energy and steam over months or years of operation, investigate different CAPEX investments, and train operators in potentially dangerous operation scenarios.

\begin{figure}[htbp]
\centering

\begin{subfigure}[b]{0.48\textwidth}
    \includegraphics[height=5cm]{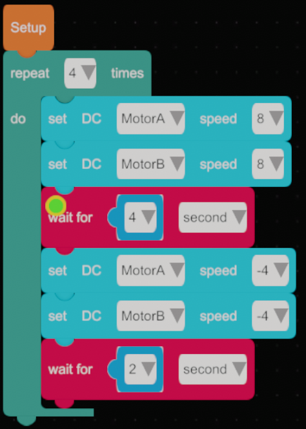}
    \caption{Example of coding interface in BRIDGE.}
    \label{fig:material3_bridge_slide1}
\end{subfigure}
\hfill
\begin{subfigure}[b]{0.48\textwidth}
    \includegraphics[height=5cm]{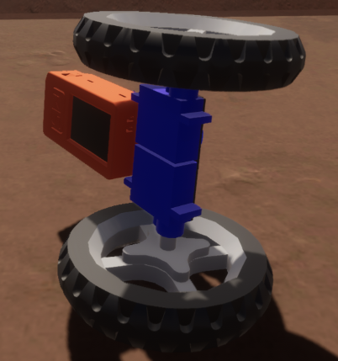}
    \caption{Flipped robot in BRIDGE.}
    \label{fig:material3_bridge_slide2}
\end{subfigure}
\hfill

\vspace{0.5cm}

\begin{subfigure}[b]{0.48\textwidth}
    \includegraphics[height=3cm]{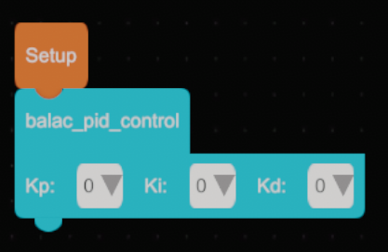}
    \caption{PID block in BRIDGE.}
    \label{fig:material3_bridge_slide4}
\end{subfigure}
\hfill
\begin{subfigure}[b]{0.48\textwidth}
    \includegraphics[height=3cm]{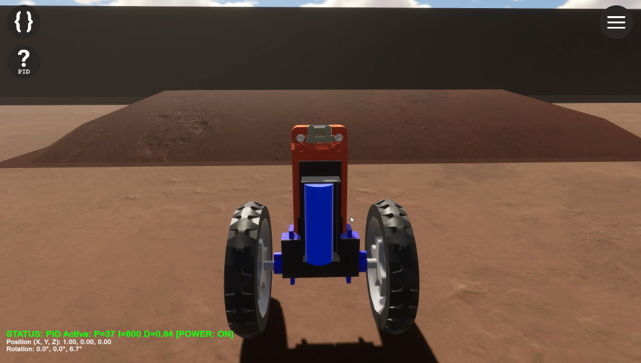}
    \caption{Balanced robot in BRIDGE.}
    \label{fig:material3_bridge_slide5}
\end{subfigure}
\hfill

\caption{Screenshots from the BRIDGE simulator.}
\label{fig:material3_bridge}
\end{figure}

After discussing the role of simulators in industry and education, students used the BRIDGE simulator to attempt balancing the robot through open-loop control by creating simple logic-based motor commands (Figure~\ref{fig:material3_bridge_slide1}). BRIDGE reproduces a virtual counterpart of the BalaC self-balancing robot used in the subsequent hands-on laboratory and combines a three-dimensional simulation environment with a block-based programming workspace. The latter mirrors the logic of the UIFlow\textsuperscript{\tiny\textregistered} environment used with the physical platform, enabling students to explore controller behaviour in simulation before moving to the real robot. This activity was designed to demonstrate the difficulty of controlling an unstable system without feedback or access to sensor measurements. Students were then challenged to intentionally flip the robot by manipulating wheel speeds and direction changes (Figure~\ref{fig:material3_bridge_slide2}), illustrating both the limitations of open-loop control and the advantages of simulation for safely exploring behaviours that could be difficult or damaging to reproduce on physical hardware. Before introducing closed-loop control, the workshop presented the standard PID controller equation (Eq.~\ref{eq:pid}) and explained the roles of the proportional, integral, and derivative gains in regulating system behaviour. The serious game at the core of the simulation step, BRIDGE, and its preliminary validation have been described in \citep{morano2026serious}

\begin{equation}
u(t)=K_p e(t)
+K_i\int_{t_1}^{t_2} e(\tau)\,d\tau
+K_d\frac{de(t)}{dt}
\label{eq:pid}
\end{equation}

The students are asked to balance the robot in BRIDGE using the "balac\_pid\_control" block, containing the logic to implement a PID controller, as a function of the parameters $K_p$, $K_i$ and $K_d$, see Figure \ref{fig:material3_bridge_slide4}. The students are encouraged to use trial-and-error to find a combination of PID parameters that work, resulting in a self-balancing robot in Figure \ref{fig:material3_bridge_slide5}.

\subsubsection{Hands-On Laboratory}

In the final part of the workshop, the students are introduced to the web-based programming interface UIFLOW\textsuperscript{\textregistered} and receive the robotic kits and hand-outs with instruction to assemble BalaC robot, to connect the M5StickC Plus microcontroller to UIFLOW\textsuperscript{\textregistered} user interface and detailed instructions to the four practical activities (make the LED light in M5StickC Plus blink; read and write sensor measurements on the M5StickC Plus display; start and stop BalaC motors; and use a custom library to balance the BalaC). The reader is referred to the M5Stack website \footnote{\url{https://m5stack.com/}} for detailed information about BalaC, UIFLOW\textsuperscript{\textregistered} and M5StickC Plus. The educational materials used in the workshops (SDT-Informed, Original and a 3-hour version in Swedish of the SDT-Informed version), the custom library for implementation and calibration of the PID controller in UiFLOW and the BRIDGE simulator are available online under a CC BY-NC-SA 4.0 license. Università Politecnica delle Marche and Optimation AB developed the original lesson materials \footnote{\url{https://zenodo.org/records/22126288}}, while ANcybernetics and Università Politecnica delle Marche jointly developed the BRIDGE serious game \footnote{\url{https://github.com/ANcybernetics/AutoSTEM-VirtualToolkit}}. The custom UIFlow\textsuperscript{\tiny\textregistered} library used for the balancing activity was also developed within AutoSTEM by Università Politecnica delle Marche and ANcybernetics. Optimation subsequently introduced the SDT-informed modifications in the workshops held in Sweden, the educational material is available in English \footnote{\url{https://doi.org/10.5281/zenodo.23124371}} and in Swedish \footnote{\url{https://doi.org/10.5281/zenodo.23124596}}.

The six workshops delivered in Sweden (Sessions 01--06) were designed
and facilitated following SDT guidelines~\citep{niemiec2009autonomy,
cheon2020autonomy, gilroy2026self}. The facilitators applied techniques based on
autonomy-supportive teaching behaviours during workshop delivery. The workshop was designed according to the three basic psychological needs proposed by Self-Determination Theory. {Autonomy} was supported by providing students with meaningful choices regarding participation, exploration of the BRIDGE simulator, and the order of laboratory activities, while facilitators used non-controlling language and encouraged self-directed learning. {Competence} was fostered through progressively scaffolded activities, informational feedback focused on effort and strategy, and an explicit emphasis on viewing mistakes as a natural and valuable part of the engineering design and PID tuning process. {Relatedness} was promoted through collaborative learning activities, paired programming, group problem solving, and active facilitator engagement with students. Facilitators sought to create an inclusive and respectful learning environment in which all contributions were valued, and female STEM professionals participated as facilitators in half of the Swedish sessions, providing visible role models and strengthening students' sense of belonging in STEM.

The Italian workshop covered the same technical content and followed the same three-lesson structure. The Swedish workshops, however, incorporated an explicitly defined SDT-informed facilitation protocol that the Italian workshop did not formally implement. In particular, the structured SDT elements adopted in Sweden, including the workshop agreements shown in \ref{fig:material1_agreements}, were not used in the Italian session. The theoretical component was delivered through a facilitator-led presentation. During the practical activities with BRIDGE and the physical robot, facilitators briefly introduced each task's objectives, after which students were given substantial operational autonomy to explore, test alternative approaches, and work on the proposed challenges, with possible solutions discussed only after the exploration phase. The Italian workshop therefore served as a comparison condition without an explicit SDT-informed facilitation protocol. The intended contrast concerns the formal, systematic use of an SDT framework, rather than the presence or absence of student autonomy or other practices potentially aligned with SDT during the practical activities. Because the facilitation condition was fully confounded with country and site, interpret this comparison as exploratory rather than causal.

\subsection{Intervention Evaluation Design, Sites and Participants}
\label{sec:meth-design}

The evaluation employs a multi-site pre-test--post-test design in which pre- and
post-questionnaire data are treated as independent cross-sectional samples (Section~\ref{sec:meth-pairing}).
Seven workshop sessions yielded data (Table~\ref{tab:sessions}): six in Sweden
(SDT-informed delivery group, $N_{\text{pre}} = 62$ students) and one in Italy
(comparison group, $N_{\text{pre}} = 19$). Participants were secondary-school
students recruited through consortium partner networks. Informed consent was obtained
from students and legal guardians.

\begin{table}[t]
\centering
\caption{Workshop session summary. SDT = Self-Determination Theory guidelines applied.}
\label{tab:sessions}
\begin{tabular}{lcccccc}
\toprule
Session & Country & Group & Dur. (h) & Female Facilitator & $N_{\text{pre}}$ & $N_{\text{post}}$ \\
\midrule
01 & Sweden & SDT & 6 & Yes & 12 & --  \\
02 & Sweden & SDT & 3 & Yes & 12 & 10  \\
03 & Sweden & SDT & 3 & Yes & 8  & 7   \\
04 & Sweden & SDT & 6 & No  & 8  & 4   \\
05 & Sweden & SDT & 3 & No  & 11 & 11  \\
06 & Sweden & SDT & 3 & No  & 11 & 9   \\
07 & Italy   & Control & 6 & Yes  & 19 & 19  \\
\midrule
\textbf{Total} & -- & -- & -- & -- & \textbf{81} & \textbf{60} \\
\bottomrule
\end{tabular}
\end{table}

\textbf{Treatment--site confounding.} The instructional contrast of interest
(SDT-informed versus comparison condition) is fully confounded with site: all SDT-informed
workshops took place in Sweden and the comparison workshop in Italy. The two
groups also differ in age composition (Swedish students: $M = 15.18$ years,
$SD = 0.59$; Italian students: $M = 16.21$, $SD = 0.42$) and in prior experience
(Section~\ref{sec:results-baseline}). Causal attribution of any between-group
difference to SDT facilitation alone is therefore not possible. Rather than ignoring
this limitation or over-interpreting the contrast, the analysis strategy described in
Section~\ref{sec:meth-analysis} (i) treats the between-condition comparison as
exploratory, (ii) controls statistically for observed individual and site-composition
differences, (iii) absorbs all stable site characteristics via session fixed effects,
and (iv) complements the between-country contrast with within-Sweden comparisons of
implementation features (duration, facilitator gender) that are free of country
confounding.

\subsection{Absence of Participant Pairing}
\label{sec:meth-pairing}

Across the Swedish sessions, the survey instruments did not include reliable
participant identifiers that would enable individual-level pre--post matching. In
several sessions, the number of pre- and post-questionnaire respondents differs (e.g.,
Session~02: 12 pre, 10 post; Session~04: 8 pre, 4 post), and there is no mechanism to
determine which pre-questionnaire respondents correspond to which post-questionnaire
respondents. Consequently, pre- and post-questionnaire data are treated as
{independent samples} drawn from the same participant pool at two time points.
This analytical approach is statistically conservative---independent-samples tests have
lower power than paired tests for the same sample size---and provides valid inferences
under the assumption that the pre- and post-samples are drawn from the same population.
All within-group analyses employ independent-samples $t$-tests with Welch's correction
for unequal variances, and effect sizes are reported as Hedges' $g$ for independent
groups. The between-condition models described in Section~\ref{sec:meth-analysis} are
estimated on the pooled pre/post records and therefore test differences in
pre--post change rather than individual change.

\subsection{Instruments and Measures}
\label{sec:meth-instruments}

The evaluation instrument is a structured pre–post questionnaire developed at Università Politecnica delle Marche as part of the AutoSTEM evaluation design and first used in the preliminary validation of BRIDGE \citep{morano2026serious}. The original instrument comprised seven perception items on robotics and control competence, three items on expectations towards the serious game, and a five-item knowledge quiz. For the present study, we added four interest items on STEM career orientation, three additional expectation items, and three open-ended questions. For the present analysis, the authors mapped the Likert items onto SCCT constructs using the structure proposed by \citet{kier2014development}.
The pre-questionnaire comprises 35 items organised into seven sections: demographics (age,
gender), background experience (prior robotics, programming, and simulator exposure),
video game usage, seven perception items (self-efficacy in control concepts), four
interest items (STEM career orientation), six expectation items (anticipated workshop
experience), and five knowledge items (multiple-choice quiz on serious games, control
systems, and PID). The post-questionnaire retains the perception, interest, and
knowledge items with identical wording; expectation items are reworded from future to
past tense. Three open-ended questions capture qualitative reflections on what worked
well, what did not work well, and what was learned. The questionnaires are available
in Appendices~\ref{app:pre} and~\ref{app:post}.

\textbf{Self-report Likert items.} Likert items are rated on a 5-point scale
(1 = Strongly disagree to 5 = Strongly agree) and are mapped to four SCCT dimensions
(Self-Efficacy, Outcome Expectations, Interest, Contextual Support) following the
validated structure of~\citet{kier2014development}. Three SDT-aligned proxy dimensions are computed as unweighted means of constituent items: Autonomy, Competence, and Relatedness. These dimensions are exploratory proxies derived retrospectively from questionnaire items that were not originally designed as validated measures of basic psychological need satisfaction. There is no consensus in the literature as to Competence and Self-efficacy being interchangeble. Some argue that the concepts are complementary in some fields (i.e. academic and athletic domains \citep{Hughes2011CompetenceSelfEfficacy}) while within SDT, Competence is viewed as the universal psychological need that in the right circumstances can lead to psychological growth, and well-being while Self-Efficacy is defined as being mechanically able to perform a task\citep{ryan2000self}. Furthermore, Autonomy in SDT is related to the person's perceived locus of causality and whatever action is driven by intrinsic motivation or integrated or identified extrinsic motivation whereas Relatedness indicates a sense of belonging, connection and significance is a community. The mapping was introduced for the purposes of the present analysis and should therefore be interpreted as an approximate alignment with SDT constructs rather than as a direct measurement of Autonomy, Competence, and Relatedness. SCCT constructs of Self-Efficacy, SE1-SE6, are directly used as proxies for Competence; SCCT constructs of Contextual Support, in particular how students perceived engagement, level of difficult and usability are used for proxies of Autonomy; and finally, SCCT constructs of Outcome Expectations and Interest, in particular how students perceive their closeness to STEM community in general and the automatic control community in particular, were used as proxies for Relatedness. Table \ref{app:table_dimension} provides the complete item-to-dimension mapping.

\textbf{Objectively scored knowledge quiz.} Five multiple-choice items with one correct
answer each (serious-game definition, purpose of a control system, meaning of the PID
acronym, components of a PID controller, definition of autonomous motion) are scored
0/1 against the correct option. The composite score is the sum of the five items
(range 0--5). Responses were scored by exact matching against answer keys in Swedish,
Italian, and English; scoring required no human judgement and is therefore objective.
Internal consistency of the binary composite was modest but acceptable for a
short knowledge test (KR-20 $= 0.58$ pre, $0.54$ post), consistent with the
multidimensional content coverage of the items.

\textbf{Qualitative items.} The three open-ended questions were coded with a manual
thematic procedure (one primary category per response; categories were derived
iteratively from the responses). Frequencies are reported in Appendix~\ref{app:open_ended}.

\subsection{Statistical Analysis}
\label{sec:meth-analysis}

All analyses were conducted in Python using \texttt{pandas}, \texttt{scipy}, and
\texttt{statsmodels}. The significance level was $\alpha = 0.05$ (two-sided
throughout). Hedges' $g$ is interpreted as: $|g| < 0.2$ negligible,
$0.2 \leq |g| < 0.5$ small, $0.5 \leq |g| < 0.8$ medium, and $|g| \geq 0.8$
large~\citep{cohen1988statistical, lakens2013calculating}.

\textbf{Outcome families and multiplicity control.} To reduce the risk of
significance-farming across a large item pool, outcomes were pre-classified into a
confirmatory family---PID self-efficacy composite, objectively scored knowledge
composite, and STEM interest composite---and a secondary family (SCCT self-efficacy,
perception, outcome expectations, contextual support, and perceived engagement). The
Holm--Bonferroni procedure controls the family-wise error rate within each family and
within each analysis (SDT and control separately).

\textbf{Within-group pre--post comparison.} For each outcome, independent-samples
$t$-tests with Welch's correction compare pre- and post-questionnaire means separately
within the SDT group and the control group, with Hedges' $g$ as the effect size and the
Mann--Whitney $U$ test as a non-parametric cross-check. Knowledge items are compared
with two-proportion $z$-tests (Holm-corrected within group) and the composite with
Welch's $t$-test; item-level percentages are accompanied by Cohen's $h$.

\textbf{Site fixed-effects models.} To absorb all stable site characteristics
(country, school, language, workshop format, facilitator, and any unmeasured site
feature that does not change between the pre- and post-waves), models of the form
$Y = \beta_0 + \beta_1 \mathrm{post} + \beta_2 (\mathrm{post} \times \mathrm{SDT})
+ \gamma^\top X + \sum_s \delta_s \mathrm{session}_s + \varepsilon$ were estimated on
the pooled pre/post records with HC1 heteroskedasticity-robust standard errors. Here
$\beta_1$ is the pre--post change in the control condition and $\beta_2$ is the
site-adjusted between-condition difference-in-differences ($\Delta\Delta$). Because
condition is constant within session, it is absorbed by the session indicators and the
contrast is identified from condition-specific within-session change.

\textbf{Multilevel models.} Hierarchical linear models with a random intercept for
session (students nested within 7 sites) estimate the same $\mathrm{post} \times
\mathrm{SDT}$ interaction and partition outcome variance into site and residual
components, with the intraclass correlation (ICC) reported as a descriptive index of
site dependence~\citep{raudenbush2002hierarchical}.

\textbf{Covariate-adjusted models.} As a specification check that does not absorb the
condition main effect, the $\Delta\Delta$ contrast was re-estimated with individual
covariates (age, gender) and session-composition covariates (session means of prior
robotics, programming, automatic-control, simulator experience, and gaming). Because
background items were administered only in the pre-questionnaire, session-composition
covariates are attached to each post-wave record from its session; this is a case-mix
adjustment at the site level. Propensity-score weighting was not pursued because with
six SDT sites and one control site the positivity assumption required for inverse
probability weighting is untenable, and extreme weights would produce unstable
estimates~\citep{rosenbaum1983central, austin2011optimal}; the covariate-adjusted and
fixed-effects models serve the same balancing purpose more transparently.

\textbf{Site heterogeneity and sensitivity.} Per-session pre--post changes are reported
with 95\% confidence intervals for the PID self-efficacy and knowledge composites.
Leave-one-session-out analysis recomputes the pooled SDT estimates with each Swedish
session removed in turn, quantifying the influence of individual sites.
Within-Sweden models interact pre--post change with workshop duration (6~h vs.\ 3~h)
and facilitator gender, with session fixed effects, providing implementation contrasts
that are entirely free of cross-country confounding.

\textbf{Gender analysis.} Gender $\times$ time interactions (female minus male
difference in pre--post change) are estimated within each condition with session fixed
effects and age as a covariate.

\textbf{Common-method diagnostics.} Harman's single-factor test was applied to all 17
Likert items (perception, interest, and expectation) separately for the pre- and
post-questionnaires. A first unrotated principal component explaining less than 50\% of
the variance is conventionally interpreted as evidence that a single common method
factor does not dominate the responses~\citep{podsakoff2003common}.

\textbf{Triangulation.} For each construct, the direction and magnitude of the
self-report, objectively scored, and qualitative evidence are tabulated side by side,
and the convergence (or divergence) of the three sources is reported explicitly.

\section{Results}
\label{sec:results}

\subsection{Workshop Delivery}
\label{sec:results-delivery}






We conducted seven workshops: six in two secondary schools in Luleå, Sweden, and one in the Marche region, Italy. To accommodate local schedules and curricula, workshops in one Swedish school lasted approximately six hours and were delivered in English, whereas workshops at the other Swedish school lasted around three hours and used a condensed Swedish translation of the material; the Italian workshop lasted approximately six hours and was delivered in Italian, with slides in English; students could choose whether to use the serious game in Italian or in English. Despite these differences, all workshops allocated substantial time to both the BRIDGE simulator and the hands-on robotics activities, with students encouraged to work collaboratively in pairs or small groups while facilitators provided guidance and support.

The Swedish workshops were delivered according to SDT principles, aiming to support students' autonomy, competence, and relatedness through collaborative, student-centred activities and a safe, inclusive learning environment. In five of the six Swedish workshops, the classroom teacher was present to support classroom management, observe the activities, and assist students when needed. The Italian workshop did not include the SDT-informed facilitation protocol: the theoretical lesson was delivered as a traditional lecture, while in the practical activities students worked freely on the tasks before solutions were presented.

Data collection involved online pre- and post-workshop questionnaires administered with the support of the participating teachers. Pre-questionnaires were generally completed one week before the intervention. Post-questionnaires were completed immediately after the workshop in Italy but at a later time in Sweden. This timing difference coincides with the condition contrast and may have contributed to between-condition differences, particularly in perceived engagement.

\subsection{Participants and Baseline Comparability}
\label{sec:results-baseline}

The total pre-questionnaire sample comprised 81 students (SDT Sweden: 62; Control
Italy: 19). In the SDT group, the gender distribution was 36 male (58\%), 23 female
(37\%), and 3 prefer-not-to-say (5\%); in the control group, 13 male (68\%), 5 female
(26\%), and 1 prefer-not-to-say (5\%). The total post-questionnaire sample comprised 60
respondents (SDT: 41; Control: 19). These are independent cross-sectional samples at
each time point, not paired observations.

Table~\ref{tab:baseline} reports the baseline comparability between groups using
standardised mean differences (SMD). Five differences are non-negligible. First,
control-group students were older than SDT-group students (16.21 vs.\ 15.18 years;
SMD $= 1.85$), reflecting the different school levels involved. Second, SDT-group
students reported more prior programming experience (80.0\% vs.\ 52.6\%; SMD $= 0.64$)
and, third, considerably more prior simulator experience (54.0\% vs.\ 0.0\%;
SMD $= 1.27$). Finally, weekday and weekend gaming both differ significantly. These
imbalances occur in {both} directions---the control group was older but had less
domain experience---which is precisely why the analyses below adjust for age, gender,
and site composition rather than relying on an unadjusted between-group comparison.

\begin{table}[t]
\centering
\caption{Baseline comparability of the SDT (Sweden) and Control (Italy) groups before the intervention (pre-questionnaire, student sample). Gaming weekdays and gaming weekends is measured between less than one and more than four hours per day, using the following classes: <1h (1), 1h-2h (2), 2h-4h (3), >4h (4). The value in the table reports the mean of the class number, from 1 to 4. Standardised mean differences (SMD) are absolute values; values $\geq 0.50$ (bold) indicate non-negligible imbalance. The sample size for itens indicated with $^*$ is reduced because one or more students did not answer the respective questionnaire item.}
\label{tab:baseline}
\begin{tabular}{lrrrrrrc}
\toprule
Item & SDT $n$ & SDT & Ctrl $n$ & Ctrl & SMD \\
\midrule
Age (years, M (SD)) & 61$^{*}$ & 15.18 (0.59) & 19 & 16.21 (0.42) & \textbf{1.85}  \\
Female participation (\%) & 62 & 37.1 & 19 & 26.3 & 0.23  \\
Prior robotics experience (\%) & 50$^*$ & 50.0 & 19 & 31.6 & 0.38  \\
Prior automatic control experience (\%) & 50$^*$ & 20.0 & 19 & 5.3 & 0.41  \\
Prior programming experience (\%) & 50$^*$ & 80.0 & 19 & 52.6 & \textbf{0.64}  \\
Prior simulator experience (\%) & 50$^*$ & 54.0 & 19 & 0.0 & \textbf{1.27}  \\
Gaming weekdays (median 1--4) & 50$^*$ & 1.5  & 19 & 2.2 & \textbf{0.60}  \\
Gaming weekends (median 1--4) & 50$^*$ & 1 & 19 & 1 & \textbf{1.09}  \\
\bottomrule

\end{tabular}
\end{table}

\subsection{Confirmatory Outcomes: Self-Reported Competence and Interest}
\label{sec:results-confirmatory}

\begin{figure}[t]
\centering
\includegraphics[width=\textwidth]{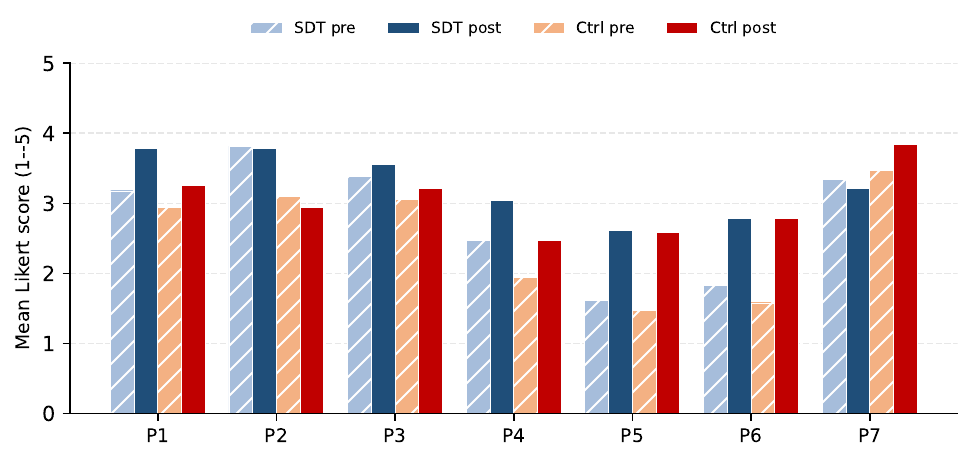}
\caption{Perception items (P1--P7) pre- and post-workshop, by condition. P5 = ``I feel
able to explain how a PID controller works''; P6 = ``I feel able to understand system
behaviour by changing P, I, D''. Error bars in the underlying analyses are omitted for
clarity; inferential results are reported in Table~\ref{tab:prepost}.}
\label{fig:perception_bars}
\end{figure}

Table~\ref{tab:prepost} reports the within-group pre--post comparisons for the
confirmatory outcomes, and Figure~\ref{fig:perception_bars} shows the item-level means
for both groups and waves.

\textbf{PID self-efficacy (large, robust gains).} The PID self-efficacy composite
(mean of the items ``I feel able to explain how a PID controller works'' and ``I feel
able to understand system behaviour by changing P, I, D'') improved from
$M = 1.72$ ($SD = 0.79$) to $M = 2.70$ in the SDT group ($\Delta M = +0.98$,
$g = 1.01$) and from $M = 1.53$ ($SD = 0.72$) to $M = 2.68$ in the control group
($\Delta M = +1.16$, $g = 1.25$). Both gains remain significant after Holm--Bonferroni
correction ($p_{\text{Holm}} < 0.001$ and $p_{\text{Holm}} = 0.001$, respectively).

\textbf{SCCT self-efficacy (medium gains).} The six-item SCCT self-efficacy composite
increased significantly in the SDT group, from $M = 2.71$ ($SD = 0.66$) to
$M = 3.26$ ($\Delta M = +0.55$, $g = 0.70$, $p_{\text{Holm}} = 0.008$), and in the
control group from $M = 2.35$ to $M = 2.88$ ($\Delta M = +0.53$, $g = 0.72$,
uncorrected $p = 0.030$; not significant after Holm correction within the secondary
family). The two largest item-level gains in both groups concerned explaining a
control system and understanding basic robotics concepts.

\textbf{Objectively scored knowledge (modest, non-significant).} The knowledge
composite (0--5) increased from $M = 3.31$ ($SD = 1.40$) to $M = 3.68$ in the SDT group
($\Delta M = +0.38$, $g = 0.27$, $p_{\text{Holm}} = 0.358$) and from $M = 2.32$ to
$M = 2.58$ in the control group ($\Delta M = +0.26$, $g = 0.20$, $p_{\text{Holm}} =
1.000$). These effects are small and do not reach significance after correction.

\textbf{STEM interest (null).} Neither condition showed significant change on the
interest composite or on any of its constituent items (automatic control, robotics,
STEM subjects, STEM career), with effect sizes close to zero (SDT $g = -0.09$;
Control $g = +0.02$). Effect sizes for the remaining secondary outcomes are reported in
Table~\ref{tab:prepost} in the Appendix.

\subsection{Objectively Scored Knowledge: Item-Level Analysis}
\label{sec:results-knowledge}

Table~\ref{tab:knowledge} reports the item-level results of the knowledge quiz.

In the SDT group, the largest improvements were observed for the serious-game
definition item (73.3\% to 85.4\%, $\Delta = +12.0$ pp) and the PID acronym (47.4\% to
58.5\%, $\Delta = +11.2$ pp). Knowledge of the PID components was essentially unchanged
(72.4\% to 70.7\%), and performance on the autonomous-motion item decreased slightly
(71.7\% to 68.3\%). In the control group, the largest gains concerned the PID
components item (31.6\% to 47.4\%, $\Delta = +15.8$ pp) and the PID acronym (21.1\% to
31.6\%, $\Delta = +10.5$ pp). After Holm correction within each condition, no
item-level change reached significance, reflecting both the modest changes and the
limited sample sizes.

Two aspects of these results are important for the interpretation of the self-report
findings. First, the objective baseline differed substantially between groups: only
about a fifth of control-group students correctly identified the PID acronym at pre-test
(21.1\%), compared with nearly half of SDT-group students (47.4\%), consistent with the
larger amount of prior programming and simulator experience in the Swedish group.
Second, both groups improved by a similar amount on the PID acronym (+11.2 and
+10.5 pp), yet the control group remained well behind in absolute terms at post-test
(58.5\% vs.\ 31.6\%).

\begin{table}[t]
\centering\small
\caption{Objectively scored knowledge quiz (multiple-choice, 0/1 items). Item-level comparisons use two-proportion $z$-tests with Holm correction within each group; the composite is analysed with Welch's $t$-test. KR-20 = 0.58 (pre), 0.54 (post). K1 = Serious-game definition, K2 = Purpose of a control system, K3 = PID acronym, K4 = PID components,K5 = Autonomous motion.}
\label{tab:knowledge}
\begin{tabular}{lrrrrrrrr}
\toprule
& \multicolumn{4}{c}{SDT (Sweden)} & \multicolumn{4}{c}{Comparison (Italy)} \\
\cmidrule(lr){2-5}\cmidrule(lr){6-9}
Item & Pre \% & Post \% & $\Delta$pp & $p_{Holm}$ & Pre \% & Post \% & $\Delta$pp & $p_{Holm}$ \\
\midrule
K1& 73.3 & 85.4 & +12.0 & 0.751 & 68.4 & 73.7 & +5.3 & 1.000 \\
K2& 83.1 & 85.4 & +2.3 & 1.000 & 63.2 & 57.9 & -5.3 & 1.000 \\
K3& 47.4 & 58.5 & +11.2 & 1.000 & 21.1 & 31.6 & +10.5 & 1.000 \\
K4& 72.4 & 70.7 & -1.7 & 1.000 & 31.6 & 47.4 & +15.8 & 1.000 \\
K5& 71.7 & 68.3 & -3.4 & 1.000 & 47.4 & 47.4 & +0.0 & 1.000 \\
\midrule
& $n$ & $M$ (SD) & $n$ & $M$ (SD) & $\Delta M$ & $g$ & $p$ & \\
\midrule
SDT & 62 & 3.31 (1.40) & 41 & 3.68 (1.37) & +0.38 & +0.27 & 0.179 \\
Comparison & 19 & 2.32 (1.49) & 19 & 2.58 (1.02) & +0.26 & +0.20 & 0.530 \\
\bottomrule
\end{tabular}
\end{table}

\begin{figure}[t]
\centering
\includegraphics[width=\textwidth]{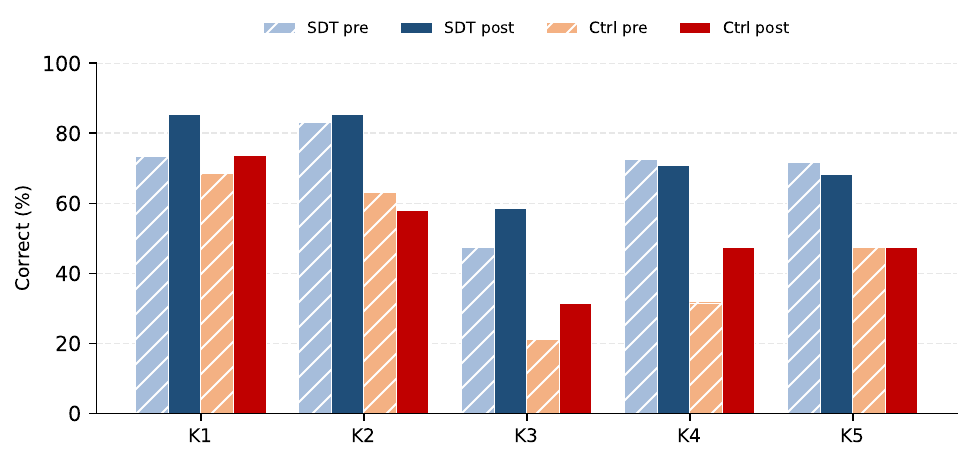}
\caption{Objectively scored knowledge quiz: percentage correct by item, condition, and
wave (K1 serious-game definition; K2 purpose of a control system; K3 PID acronym;
K4 PID components; K5 autonomous motion).}
\label{fig:knowledge_bars}
\end{figure}

\subsection{Site-Adjusted Between-Condition Comparisons}
\label{sec:results-adjusted}

\begin{figure}[t]
\centering
\includegraphics[width=\textwidth]{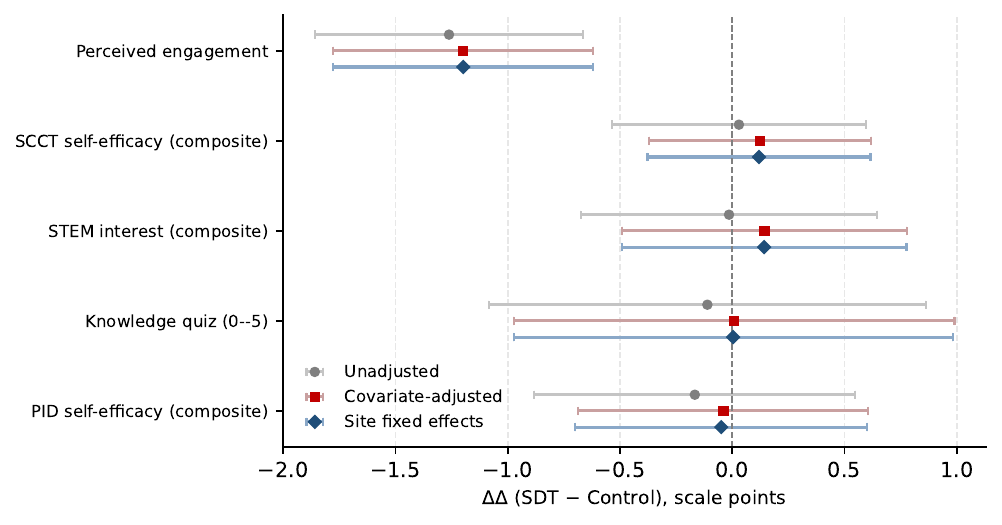}
\caption{Between-condition difference-in-differences ($\Delta\Delta$ = SDT minus
Comparison) for the confirmatory outcomes, the SCCT self-efficacy composite, and perceived
engagement under three specifications: unadjusted, covariate-adjusted, and site fixed
effects. Primary-outcome contrasts are close to zero and non-significant; engagement
favours the comparison group under every specification. The site fixed-effects estimates
are effectively identical to the covariate-adjusted estimates.}
\label{fig:adjusted}
\end{figure}

Table~\ref{tab:fe} reports the site fixed-effects models, Table~\ref{tab:mlm} the
multilevel models, and Table~\ref{tab:adjusted} the unadjusted, covariate-adjusted, and
site fixed-effects $\Delta\Delta$ contrasts side by side. Figure~\ref{fig:adjusted}
displays the three specifications graphically.

\begin{itemize}
  \item \textbf{PID self-efficacy.} Unadjusted $\Delta\Delta = -0.17$; covariate-adjusted
  $\Delta\Delta = -0.04$ (95\% CI $[-0.69, +0.61]$, $p = 0.905$); site fixed effects
  $\Delta\Delta = -0.05$ ($p = 0.883$); multilevel $\Delta\Delta = -0.04$ ($p = 0.908$).
  \item \textbf{Knowledge quiz.} Unadjusted $\Delta\Delta = -0.11$; covariate-adjusted
  $\Delta\Delta = +0.01$ ($p = 0.987$); site fixed effects $\Delta\Delta = +0.00$
  ($p = 0.994$); multilevel $\Delta\Delta = +0.04$ ($p = 0.934$).
  \item \textbf{STEM interest.} Covariate-adjusted $\Delta\Delta = +0.14$ ($p = 0.657$);
  site fixed effects $\Delta\Delta = +0.14$ ($p = 0.660$).
  \item \textbf{SCCT self-efficacy.} Covariate-adjusted $\Delta\Delta = +0.12$
  ($p = 0.622$); site fixed effects $\Delta\Delta = +0.12$ ($p = 0.638$).
  \item \textbf{Perceived engagement.} Unadjusted $\Delta\Delta = -1.26$;
  covariate-adjusted $\Delta\Delta = -1.20$ (95\% CI $[-1.78, -0.62]$, $p < 0.001$);
  site fixed effects $\Delta\Delta = -1.20$ ($p < 0.001$); multilevel
  $\Delta\Delta = -1.21$ ($p < 0.001$).
\end{itemize}

No site-adjusted between-condition contrast on the confirmatory outcomes is
statistically significant, and all point estimates are small relative to the
within-group changes reported in Section~\ref{sec:results-confirmatory}. The single
robust adjusted difference concerns perceived engagement, which improved more in the
control group; because the engagement item changed wording from expectations
(pre) to experience (post) and baseline expectations differed substantially between
groups, this contrast is interpreted cautiously in Section~\ref{sec:disc-interest}.
The site fixed-effects models show a large, highly significant pre--post change in
the comparison condition for PID self-efficacy ($\beta = +1.16$, $SE = 0.27$,
$p < 0.001$) and a significant change for SCCT self-efficacy ($\beta = +0.53$,
$SE = 0.20$, $p = 0.009$). The multilevel models indicate modest site dependence,
which is strongest for the knowledge composite (ICC $= 0.22$) and STEM interest
(ICC $= 0.19$) and weak for PID and SCCT self-efficacy (ICC $= 0.06$ and $0.05$,
respectively). In other words, once stable site characteristics are absorbed, the
data provide no evidence that SDT-informed facilitation produced larger gains than
standard delivery of the same technical content; they also provide no evidence of
harm. The between-condition difference is estimated with sufficient precision to rule
out differences larger than roughly 0.6 scale points for PID self-efficacy
(95\% CI $[-0.69, +0.61]$).

\begin{table}[t]
\centering\small
\caption{Site fixed-effects models (session fixed effects, HC1 robust standard errors). ``Change'' is the pre--post change in the comparison condition; ``$\Delta\Delta$'' is the additional change in the SDT condition, i.e.\ the site-adjusted between-condition contrast. Session fixed effects absorb all stable site characteristics (country, school, language, workshop format).}
\label{tab:fe}
\begin{tabular}{lrrrr}
\toprule
Outcome & Change (Ctrl) & $\Delta\Delta$ (SDT$-$Ctrl) & $p$ & $n$ \\
\midrule
PID self-efficacy (composite) & +1.16 (0.27) & -0.05 (0.33) & 0.883 & 140 \\
Knowledge quiz (0--5) & +0.26 (0.43) & +0.00 (0.50) & 0.994 & 140 \\
STEM interest (composite) & +0.01 (0.24) & +0.14 (0.32) & 0.660 & 140 \\
SCCT self-efficacy (composite) & +0.53 (0.20) & +0.12 (0.25) & 0.638 & 140 \\
Perceived engagement & +0.89 (0.22) & -1.20 (0.30) & $<0.001$$^{***}$ & 127 \\
\bottomrule
\end{tabular}
\end{table}

\begin{table}[t]
\centering\small
\caption{Multilevel models with random intercepts for workshop session (students nested in 7 sessions; 5 sessions with available post-questionnaire data contribute to the interaction, post=questionnaire data from 2 sessions are not available). $\Delta\Delta$ = post $\times$ SDT interaction; $\sigma^2_{site}$ and $\sigma^2_{resid}$ are the session and residual variance components; ICC = intraclass correlation.}
\label{tab:mlm}
\begin{tabular}{lrrrrrr}
\toprule
Outcome & $\Delta\Delta$ (SE) & $p$ & $\sigma^2_{site}$ & $\sigma^2_{resid}$ & ICC & $n$ \\
\midrule
PID self-efficacy (composite) & -0.04 (0.34) & 0.908 & 0.050 & 0.771 & 0.06 & 140 \\
Knowledge quiz (0--5) & +0.04 (0.49) & 0.934 & 0.464 & 1.611 & 0.22 & 140 \\
STEM interest (composite) & +0.10 (0.35) & 0.768 & 0.191 & 0.823 & 0.19 & 140 \\
SCCT self-efficacy (composite) & +0.14 (0.27) & 0.603 & 0.023 & 0.478 & 0.05 & 140 \\
Perceived engagement & -1.21 (0.34) & $<0.001$$^{***}$ & 0.137 & 0.749 & 0.15 & 127 \\
\bottomrule
\end{tabular}
\end{table}

\begin{table}[t]
\centering\small
\caption{Between-condition difference-in-differences ($\Delta\Delta$ = SDT minus Comparison change) under three specifications. Unadjusted: pooled pre/post OLS. Covariate-adjusted: adds age, gender and session-composition case-mix covariates (prior robotics, programming, automatic control, simulator use and gaming). Site fixed effects: session dummies absorb all stable site differences, so $\Delta\Delta$ is identified from condition-specific pre--post change. Robust (HC1) standard errors. $^{*}p<0.05$, $^{**}p<0.01$, $^{***}p<0.001$.}
\label{tab:adjusted}
\begin{tabular}{lrrrrr}
\toprule
Outcome & Unadj. & Covariate-adj. [95\% CI] & $p$ & Site FE [95\% CI] & $p$ \\
\midrule
PID self-efficacy & -0.17 & -0.04 [-0.69, +0.61] & 0.905 & -0.05 [-0.70, +0.60] & 0.883 \\
Knowledge quiz & -0.11 & +0.01 [-0.97, +0.99] & 0.987 & +0.00 [-0.97, +0.98] & 0.994 \\
STEM interest & -0.01 & +0.14 [-0.49, +0.78] & 0.657 & +0.14 [-0.49, +0.78] & 0.660 \\
SCCT self-efficacy & +0.03 & +0.12 [-0.37, +0.62] & 0.622 & +0.12 [-0.38, +0.62] & 0.638 \\
Perceived engagement & -1.26 & -1.20 [-1.78, -0.62] & $<0.001$$^{***}$ & -1.20 [-1.78, -0.62] & $<0.001$$^{***}$ \\
\bottomrule
\end{tabular}
\end{table}

\subsection{Site Heterogeneity and Sensitivity Analyses}
\label{sec:results-sensitivity}

\begin{figure}[t]
\centering
\includegraphics[width=\textwidth]{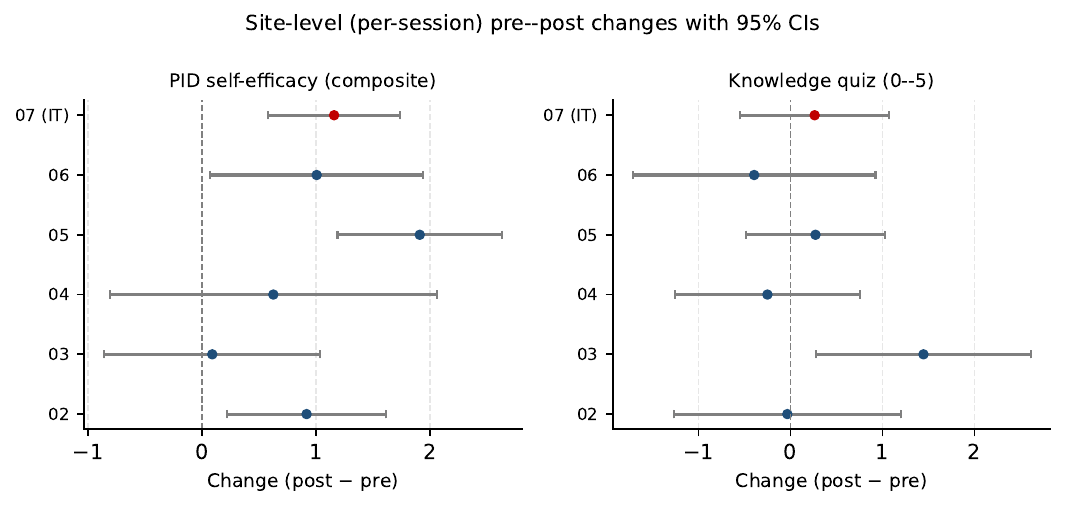}
\caption{Per-session pre--post changes with 95\% confidence intervals for PID
self-efficacy (left) and the objectively scored knowledge composite (right). Session 07
is the Italian comparison workshop. PID gains are positive in every session; knowledge
changes are heterogeneous.}
\label{fig:session_effects}
\end{figure}

\textbf{Per-session changes.} Figure~\ref{fig:session_effects} shows the pre--post
change in PID self-efficacy and the knowledge composite for each session. PID
self-efficacy gains are positive in all six sessions with post-data, including the
Italian comparison session, and the confidence intervals exclude zero in the two largest
Swedish sessions and in Italy. Knowledge changes are small and heterogeneous in sign
across sessions, consistent with the null composite result.

\begin{figure}[t]
\centering
\includegraphics[width=\textwidth]{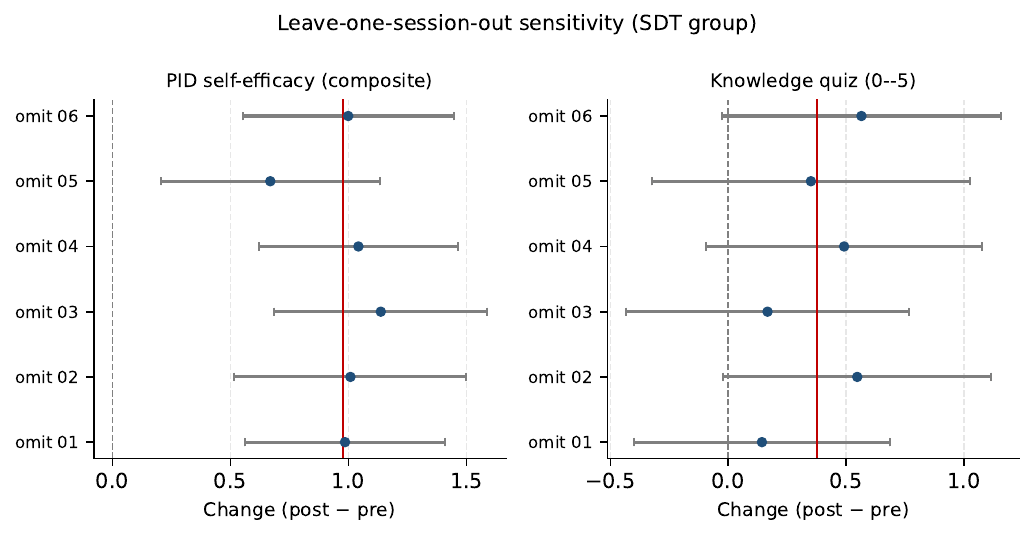}
\caption{Leave-one-session-out sensitivity analysis for the SDT group. Vertical red
lines mark the pooled estimate; all PID self-efficacy estimates remain positive and
significant.}
\label{fig:l1so}
\end{figure}

\textbf{Leave-one-session-out.} Table~\ref{tab:l1so} and Figure~\ref{fig:l1so} report
the pooled SDT estimates recomputed with each Swedish session removed in turn. The PID
self-efficacy gain remains positive and significant in every specification, ranging
from $\Delta M = +0.67$ ($g = 0.70$, $p = 0.007$, when session~05 is removed) to
$\Delta M = +1.14$ ($g = 1.18$), and the SCCT self-efficacy gain ranges from $+0.34$
to $+0.62$ scale points. The knowledge composite ranges from $+0.14$ to $+0.57$ and
remains non-significant throughout. No single site drives the self-efficacy findings.

\begin{table}[t]
\centering\small
\caption{Leave-one-session-out sensitivity analysis for the SDT (Sweden) group: $\Delta$ and Hedges' $g$ recomputed with each Swedish session removed in turn.}
\label{tab:l1so}
\begin{tabular}{lrrr}
\toprule
Sample & $\Delta M$ & Hedges' $g$ & $p$ \\
\midrule
\multicolumn{4}{l}{\textit{PID self-efficacy (composite)}} \\[1pt]
\quad All sessions & +0.98 & +1.01 & $<0.001$$^{***}$ \\
\quad omit session 01 & +0.99 & +0.99 & $<0.001$$^{***}$ \\
\quad omit session 02 & +1.01 & +0.99 & $<0.001$$^{***}$ \\
\quad omit session 03 & +1.14 & +1.18 & $<0.001$$^{***}$ \\
\quad omit session 04 & +1.04 & +1.14 & $<0.001$$^{***}$ \\
\quad omit session 05 & +0.67 & +0.70 & 0.007$^{**}$ \\
\quad omit session 06 & +1.00 & +1.05 & $<0.001$$^{***}$ \\
\multicolumn{4}{l}{\textit{Knowledge quiz (0--5)}} \\[1pt]
\quad All sessions & +0.38 & +0.27 & 0.179 \\
\quad omit session 01 & +0.14 & +0.11 & 0.608 \\
\quad omit session 02 & +0.55 & +0.41 & 0.063 \\
\quad omit session 03 & +0.17 & +0.12 & 0.588 \\
\quad omit session 04 & +0.49 & +0.35 & 0.104 \\
\quad omit session 05 & +0.35 & +0.23 & 0.314 \\
\quad omit session 06 & +0.57 & +0.41 & 0.065 \\
\multicolumn{4}{l}{\textit{SCCT self-efficacy (composite)}} \\[1pt]
\quad All sessions & +0.55 & +0.70 & 0.002$^{**}$ \\
\quad omit session 01 & +0.56 & +0.68 & 0.002$^{**}$ \\
\quad omit session 02 & +0.58 & +0.72 & 0.004$^{**}$ \\
\quad omit session 03 & +0.56 & +0.69 & 0.005$^{**}$ \\
\quad omit session 04 & +0.62 & +0.82 & $<0.001$$^{***}$ \\
\quad omit session 05 & +0.34 & +0.42 & 0.091 \\
\quad omit session 06 & +0.62 & +0.86 & $<0.001$$^{***}$ \\
\bottomrule
\end{tabular}
\end{table}

\textbf{Within-Sweden implementation moderators.} Table~\ref{tab:moderators} reports
pre-post change interacted with workshop duration and facilitator gender, estimated
with session fixed effects. None of the contrasts is significant at
$\alpha = 0.05$. Descriptively, the largest coefficient is a $-0.74$ point difference
($p = 0.070$) in PID self-efficacy change between sessions with a female versus a male
facilitator, and a $+0.87$ point difference ($p = 0.120$) in knowledge change for the
same contrast; these estimates should be treated as hypothesis-generating given the
small number of sessions. Duration (6 h vs. 3 h) shows no meaningful association with any outcome, although this contrast should be interpreted cautiously given the limited number of sessions contributing post-workshop data and the overlap between duration, school, and language.

\begin{table}[t]
\centering\small
\caption{Within-Sweden moderator analyses (sessions 01--06): pre--post change interacted with workshop duration (6\,h vs.\ 3\,h) and facilitator gender, with session fixed effects and individual covariates. These contrasts are free of cross-country confounding.}
\label{tab:moderators}
\begin{tabular}{lrrrr}
\toprule
Outcome & 6\,h vs.\ 3\,h & $p$ & Female vs.\ male fac. & $p$ \\
\midrule
PID self-efficacy & -0.45 & 0.533 & -0.74 & 0.070 \\
Knowledge quiz & +0.24 & 0.709 & +0.87 & 0.120 \\
SCCT self-efficacy  & -0.34 & 0.551 & -0.10 & 0.761 \\
\bottomrule
\end{tabular}
\end{table}

\subsection{Gender Differences in Change}
\label{sec:results-gender}

Table~\ref{tab:gender} reports gender $\times$ time interactions with session fixed
effects. Female participants gained substantially less than male participants on PID
self-efficacy in the SDT condition ($-1.06$ scale points, 95\% CI $[-1.80, -0.31]$,
$p = 0.006$) and a similar, marginally non-significant deficit in the comparison condition
($-0.92$, 95\% CI $[-1.87, +0.04]$, $p = 0.060$). Knowledge gains did not differ
significantly by gender in either condition.

Because the comparison group contained only five female respondents, the comparison-group estimate is imprecise; nevertheless, the magnitude of the female deficit is similar in both conditions, which argues against interpreting the gap as a unique consequence of
SDT-informed facilitation. It is more plausibly related to broader gendered patterns in
technical confidence and participation---patterns that the intervention, as delivered,
did not neutralise. Second, because pre- and post-samples are independent, the
interaction estimates capture differences between the female and male \emph{samples}
at the two waves rather than within-person change, and they should be read with the
caution appropriate to unpaired data.

\begin{table}[t]
\centering\small
\caption{Gender $\times$ time interactions with session fixed effects: the coefficient is the additional pre--post change for female relative to male participants (female $-$ male gap in change), adjusting for age and site. Robust (HC1) standard errors. $^{*}p<0.05$, $^{**}p<0.01$, $^{***}p<0.001$.}
\label{tab:gender}
\begin{tabular}{lrrrr}
\toprule
Outcome & Female $-$ male $\Delta$ [95\% CI] & $p$ & $n_F$ (post) & $n$ \\
\midrule
\multicolumn{5}{l}{\textbf{SDT (Sweden)}} \\[1pt]
\quad PID self-efficacy& -1.06 [-1.80, -0.31] & 0.006$^{**}$ & 20 & 102 \\
\quad Knowledge quiz & -0.48 [-1.48, +0.52] & 0.348 & 20 & 102 \\
\multicolumn{5}{l}{\textbf{Comparison (Italy)}} \\[1pt]
\quad PID self-efficacy & -0.92 [-1.87, +0.04] & 0.060 & 5 & 38 \\
\quad Knowledge quiz & +1.10 [-0.53, +2.73] & 0.185 & 5 & 38 \\
\bottomrule
\end{tabular}
\end{table}

\subsection{Qualitative Feedback from Open-Ended Responses}
\label{sec:open_ended_results}

Qualitative feedback was collected through three open-ended post-workshop questions
concerning positive aspects of the workshop, challenges encountered, and learning
outcomes. Of the 41 Swedish respondents, 40 provided at least one substantive response, whereas the open-ended questions were not part of the Italian questionnaire, which had been administered before these items were added to the instrument. Responses were manually coded, and
detailed frequency distributions are presented in Appendix~\ref{app:open_ended}
(Tables~C.1--C.3). Hands-on activities involving the robot and simulator emerged as the
most frequently cited strengths (22\% and 10\% of respondents, respectively), alongside
generally positive evaluations of the workshop, collaboration opportunities, clear
exercise instructions, and facilitator support.

The most commonly reported challenges were technical issues (20\%), including robot
connection problems, simulator platform incompatibility, and hardware fragility, and
the length of the presentations (12\%). Several students also commented on
difficulties understanding some PID-related concepts and instructions; at the same
time, 20\% of respondents explicitly reported that nothing had gone wrong. Regarding
learning outcomes, students most frequently reported gaining knowledge about PID
controllers and automatic control (17\%), programming (15\%), and robot balancing
(10\%). These qualitative findings are consistent with the quantitative results, where
the largest improvements were observed for self-efficacy related to explaining PID
controllers and adjusting PID parameters.

\subsection{Triangulation and Common-Method Diagnostics}
\label{sec:results-triangulation}

Table~\ref{tab:triangulation} places the three evidence sources side by side for each
construct. For PID controller competence, self-report gains (SDT $g = 1.01$; Control
$g = 1.25$), qualitative accounts (17\% of respondents cited PID/control concepts as
their main learning outcome), and the objective PID-acronym item ($+11.2$ and $+10.5$
pp) all point in the same direction, although the objective PID-components item was
flat in the SDT group and improved in the comparison group. For control-system
understanding, self-report gains ($g = 0.55$ and $0.57$) are accompanied by a flat or
negative objective item and by qualitative themes emphasising programming and robot
balancing rather than control theory per se. Basic
robotics shows convergence between self-report and qualitative evidence but only
partial agreement with the objective items. STEM interest shows a convergent null
across all three sources.

Furthermore, the Harman's single-factor test indicates that the first unrotated principal
component accounted for 33.1\% of the variance in the pre-questionnaire and 45.2\% in
the post-questionnaire, below the 50\% heuristic, indicating that a single common method
factor does not dominate the responses. Combined with the divergence observed between
self-report and objective knowledge items, this suggests that the self-report gains are
not merely an artefact of method variance, while also cautioning against treating
self-reported and objectively scored outcomes as interchangeable.

\begin{table}[t]
\centering
\small
\caption{Triangulation of evidence across the three data sources for the Swedish workshops ($N=41$). Qualitative evidence is available for the Swedish workshops only, while self-report and knowledge results cover both conditions.}
\label{tab:triangulation}

\begin{tabular}{p{3.0cm}p{2.5cm}p{2.5cm}p{4.3cm}}
\toprule
\textbf{Construct} &
\textbf{Self-report} &
\textbf{Objective knowledge} &
\textbf{Qualitative evidence} \\
\midrule

\textbf{PID controller competence}
&
SDT: $g=1.01$\newline Ctrl: $g=1.25$
&
K3: $+11/+11$ pp\newline K4: $-2/+16$ pp
&
PID/control concepts cited by 7/41 (17\%) as a learning outcome
\\[0.25cm]

\textbf{Control-system understanding}
&
SDT: $g=0.55$\newline Ctrl: $g=0.57$
&
K2: $+2/-5$ pp
&
Programming cited by 6/41 (15\%); robot balancing by 4/41 (10\%)
\\[0.25cm]

\textbf{Basic robotics}
&
SDT: $g=0.53$\newline Ctrl: $g=0.32$
&
K5: $-3/+0$ pp\newline K1: $+12/+5$ pp
&
Practical/hands-on cited by 9/41 (22\%) as the strongest element
\\[0.25cm]

\textbf{STEM interest and career plans}
&
SDT: $g=-0.09$ (ns)\newline Ctrl: $g=0.02$ (ns)
&
No corresponding knowledge item
&
Occasional disinterest (1/41)
\\

\bottomrule
\end{tabular}
\end{table}


\section{Discussion}
\label{sec:discussion}

\subsection{Main Findings}
\label{sec:disc-principal}

Four findings emerge from this evaluation. First, the AutoSTEM workshop produced large,
robust gains in PID-specific self-efficacy in both conditions ($g = 1.01$ and $1.25$),
which survived multiplicity correction and every leave-one-session-out specification.
Second, the objectively scored knowledge composite improved only modestly and
non-significantly, and the self-report and objective evidence converged most clearly for
PID concept recognition rather than for conceptual control knowledge. Third, after
adjusting for site-specific differences with three complementary specifications, no
contrast on any competence or knowledge outcome was significant: SDT-informed
facilitation did not produce detectably larger gains than the comparison condition without an explicit SDT-informed facilitation protocol of the
same technical content (the one robust adjusted contrast, perceived engagement,
favoured the comparison group and is attributable to baseline-expectation and item-wording
differences). Fourth, female participants reported smaller PID self-efficacy gains than
males in both conditions, indicating a participation or confidence dynamic that was not
specific to the SDT condition. These findings are further detailed below.

\subsubsection{Competence Gains Across Instructional Conditions}
\label{sec:disc-pid}

The most prominent finding is the large and statistically significant improvement in
PID-specific self-efficacy observed in both instructional conditions. Students entered
the workshops with very low confidence in their ability to explain PID controllers
($M \approx 1.5$--$1.7$ in both groups) and left with substantially higher confidence
($M \approx 2.7$). Effect sizes are classified as large by conventional standards and
are consistent with---and in several cases exceed---the meta-analytic benchmark of
$g = 0.786$ for robotics education interventions~\citep{urrea2025industrial}.

The absence of a site-adjusted advantage for SDT-informed facilitation suggests that the core instructional components---the BRIDGE simulator’s safe-to-fail experimentation environment, the hands-on BalaC hardware manipulation, and the structured progression from theory to simulation to physical implementation---may contribute substantially to the observed gains across both instructional conditions, irrespective of the additional explicit SDT-informed facilitation strategies used in the Swedish workshops. These findings broadly align with previous educational robotics research showing that different instructional approaches can support learning when students engage in hands-on, technically authentic activities \citep{andic2024direct}, although the present study does not directly compare direct and indirect instruction.

From an SDT perspective, the BRIDGE simulator itself may intrinsically support
competence through its progressive challenge structure and immediate visual feedback,
effectively providing competence support even in the absence of facilitator-level SDT
strategies. This interpretation aligns with \citet{gilroy2026self}'s ``intertwined
pillars'' finding of a feedback cycle between competence and autonomy, and with
findings by \citet{farrell_applying_2014} that well-designed serious games can satisfy
basic psychological needs through their structural features, independent of facilitator
behaviour. It also matches \citet{nunez2025how}'s finding that the effects of autonomy
support are contemporaneous rather than cumulative---in a single-session workshop, the
immediate competence feedback from the technology may dominate the more diffuse effect
of facilitator behaviour. An important caveat is statistical: with six SDT sessions and
one comparison session, the between-condition contrast has limited power, and the
confidence intervals from the site-adjusted models (e.g., $[-0.69, +0.61]$ for PID
self-efficacy) cannot exclude modest effects in either direction. The correct
conclusion is therefore not that SDT-informed facilitation is ineffective, but that
this design cannot detect its incremental contribution over an already
competence-supportive technology-rich activity.

\subsubsection{Objective Knowledge, Self-Reports, and What Each Measures}
\label{sec:disc-objective}

The divergence between the large self-report gains and the modest objective knowledge
gains deserves attention. Self-efficacy and knowledge are conceptually distinct
outcomes: a student may become substantially more confident in explaining PID control
and in experimenting with parameters after successfully balancing a robot, while still
missing, for example, the formal composition of a PID controller on a multiple-choice
item. The triangulation shows convergence for the PID acronym (both groups improved by
10--11 pp), which is directly taught and practised, and weaker convergence for more
conceptual items. This pattern is consistent with the broader literature:
self-reports of competence correlate only moderately with performance-based
measures~\citep{duckworth2015measurement}, and conclusions about learning should rest on
multiple indicators. The present study's contribution is to report both, rather than
privileging the more flattering self-report result.

It is also important to distinguish \emph{statistical} from \emph{educational}
significance. The knowledge composite confidence intervals are wide (e.g.,
$\Delta\Delta = +0.01$, 95\% CI $[-0.97, +0.99]$), and the KR-20 for a five-item quiz
limits precision. The quiz should be understood as one anchor of a multi-method
strategy, not as a definitive measure of conceptual learning; future implementations
should add longer knowledge assessments, performance tasks (e.g., structured
PID-tuning tasks with logged outcomes), and repeated measurement occasions.

\subsubsection{STEM Interest and Engagement}
\label{sec:disc-interest}

Neither instructional condition produced significant changes in STEM career interest or
aspirations, replicating a well-established pattern: domain-specific self-efficacy is
malleable following mastery experiences, whereas career dispositions require sustained
exposure~\citep{kier2014development, wang2023gender}. From an SDT perspective, a single
workshop---however need-supportive---is unlikely to internalise STEM interest to the
point of measurable career-orientation shift; cumulative autonomy support over weeks
and months is necessary~\citep{patall2021daily, leon2015self}, which complements the
recommendation that interest-shifting interventions should employ multi-session,
longitudinal designs~\citep{dresser2024increasing, hernandez2024challenge}.

One between-group finding was statistically robust: the comparison group reported larger
engagement gains than the SDT group ($\Delta\Delta = -1.20$, 95\% CI $[-1.78, -0.62]$,
$p < 0.001$ in the site fixed-effects and covariate-adjusted models; see
Tables~\ref{tab:fe} and~\ref{tab:adjusted}). This contrast requires careful
interpretation. The engagement item was reworded from an expectation before the
workshop (``I think the activities will be engaging'') to an experience afterwards
(``I think the activities were engaging''), and the SDT group entered with markedly
higher expectations (3.69 vs.\ 3.37 on a 5-point scale). Lower expectations leave
more room for positive disconfirmation, and the comparison group's post-workshop rating
(4.26) was indeed higher than the SDT group's (3.34), while both remained above the
scale midpoint. Age composition and language also differed between groups. The
contrast is therefore best read as a difference in expectation calibration and
sample composition rather than as evidence that the comparison condition is intrinsically
more engaging than SDT-informed delivery; disentangling the two would require
identical engagement measures at both waves and within-site randomisation.

\subsubsection{Gender Dynamics}
\label{sec:disc-gender}

The gender $\times$ time analysis revealed that female participants reported smaller
PID self-efficacy gains than males in the SDT condition ($-1.06$ scale points,
$p = 0.006$), with a similar, marginally non-significant deficit in the comparison
condition ($-0.92$, $p = 0.060$). Two interpretations are plausible: first, documented gendered participation patterns in technical activities
(boys taking control of hardware and simulation tools) can translate into unequal
competence experiences even within autonomy-supportive settings; second, female
students are known to report lower confidence despite equivalent performance in STEM,
so part of the gap may reflect differential reporting rather than differential
learning. The objective knowledge quiz showed no significant gender differences,
which is more consistent with the second interpretation, but the small female
subgroups do not allow for conclusions. Importantly, the gap was not specific to the SDT
condition. Structured participation protocols---turn-taking for
simulator and hardware access, role rotation within groups, and active facilitator
monitoring of participation equity---are recommended as complements to autonomy
support~\citep{hernandez2024challenge, kovaleva2024pros}, and future studies should
combine performance measures with repeated confidence measures to disentangle
reporting from learning.

\subsection{Methodological Implications and Limitations}
\label{sec:disc-methods}

This study illustrates a practical approach to the two methodological weaknesses that
motivated it. For self-report reliance, three inexpensive steps proved informative:
(i) embedding an objectively scored quiz in the same instrument, (ii) reporting its
results alongside perceptions with explicit attention to convergence and divergence,
and (iii) reporting a common-method diagnostic. The triangulation table
(Table~\ref{tab:triangulation}) shows how much interpretive value comes from simply
placing evidence sources side by side; it also prevents the strongest self-report
effect from being reported as if it were the whole story.

For site confounding, the layered strategy employed here---multilevel variance
partitioning, session fixed effects that absorb all stable site characteristics, and
within-country implementation contrasts---is feasible even with small numbers of
sites, provided that the result is presented as an adjusted, exploratory contrast
rather than a definitive causal test. Two design lessons follow. First, future
multi-site studies should preferably randomise the instructional condition
{within} sites (or at least include more than one comparison site), which would
break the treatment--site confound at source. Second, the field would benefit from
pre-registered analysis plans that specify the primary outcome family and the
site-adjustment strategy before data collection
\citep{shadish2002experimental, lakens2013calculating}.

Several limitations constrain the interpretation of these findings.
\begin{enumerate}
    \item The study's quasi-experimental design confounds instructional approach with
    country, workshop duration, facilitator characteristics, student age, and cultural
    context. Baseline comparisons (Table~\ref{tab:baseline}) revealed substantial
    differences in age and prior experience. Although the site-adjusted models control
    for observed differences and absorb all stable site characteristics, they cannot
    rule out site-specific time trends, and the between-condition contrast remains
    exploratory.

    \item The study has only one comparison site. This limits both statistical power for
    the between-condition contrast and the generalisability of the comparison condition;
    the multilevel variance components and ICCs should be interpreted with this in mind.

    \item The knowledge quiz is short (five items; KR-20 of 0.54--0.58), which limits
    both reliability and content coverage. It anchors the multi-method strategy but
    does not substitute for a comprehensive conceptual assessment.

    \item Small sample sizes in several workshop sessions, particularly within the
    Swedish group, limited statistical power and prevented the use of more advanced
    multivariate analyses capable of controlling simultaneously for many demographic
    and background variables.

    \item The SDT-related measures of autonomy, competence, and relatedness were
    derived from questionnaire items originally designed for SCCT constructs and should
    therefore be interpreted as exploratory indicators rather than validated measures of
    basic psychological need satisfaction. This limitation reflects the broader lack of
    domain-specific SDT instruments for informal STEM learning contexts identified by
    \citet{gilroy2026self}. Future studies should employ established SDT instruments
    such as those proposed by \citet{vansteenkiste2020basic}, \citet{niemiec2009autonomy},
    and \citet{leon2015self}.

    \item The use of unpaired statistical analyses reduced analytical efficiency and
    likely attenuated the reported effect sizes compared to what could have been
    achieved with a paired pre--post design. Future implementations should employ
    persistent anonymous participant identifiers to enable more robust longitudinal
    analyses, including individual-level covariates and change-score modelling.

    \item Open-ended questions were added to the post-questionnaire after the Italian workshop and are therefore available only for the Swedish sessions, so the qualitative evidence cannot be compared across conditions.

    \item Response attrition differed across sessions, and because individual-level pairing was not available for the Swedish workshops, changes in sample composition between pre- and post-assessment cannot be excluded.
    
    \item No formal implementation-fidelity measure was collected to verify the consistency with which the SDT-informed facilitation strategies were applied across the Swedish workshops; therefore, variability in implementation across sessions cannot be ruled out.
\end{enumerate}

\section{Conclusions}
\label{sec:conclusion}

The results of the AutoSTEM project indicate that a simulation-to-real robotics workshop combining the BRIDGE serious game with tangible PID-controlled hardware (BalaC + M5StickC Plus) was associated with large gains in secondary students’ PID self-efficacy ($g = 1.01$--$1.25$), surviving Holm–Bonferroni correction and every leave-one-session-out specification, with positive gains observed across the different instructional conditions and implementation contexts considered in this study. The gains are corroborated by the objective
knowledge evidence for PID concept recognition and by qualitative accounts of learning,
while the objectively scored knowledge composite as a whole shows only small,
non-significant improvement.

After controlling for site-specific differences with session fixed effects, multilevel
models, and covariate adjustment, no between-condition contrast on any competence or
knowledge outcome was significant (the only robust difference, perceived engagement, favoured the comparison group and may partly reflect baseline-expectation and item-wording differences;
Section~\ref{sec:disc-interest}). This suggests that the AutoSTEM intervention, including its technology-supported activities---such as iterative experimentation, immediate system feedback, and safe-to-fail exploration---may itself substantially support competence development, consistent with the
``intertwined pillars'' dynamics of \citet{gilroy2026self}. This conclusion applies to
the incremental effect of SDT-informed facilitation over and above an already
competence-supportive technology, and it is bounded by the single comparison site
available in this evaluation.

Three qualifications warrant attention. First, female participants reported smaller
PID gains than males in both conditions, indicating that confidence and participation
equity require explicit structuring rather than emerging automatically from
autonomy-supportive design. Second, null findings on STEM career interest confirm that
single-session interventions are insufficient to shift durable career dispositions;
sustained, multi-session programming is needed. Third, the study's reliance on
independent samples---a valid, conservative approach when participant identifiers are
unavailable---should be considered when comparing effect sizes with paired-design
benchmarks~\citep{urrea2025industrial}.

Methodologically, the study shows that a multi-method, site-adjusted evaluation is
feasible for small-scale interventions: objectively scored items, qualitative coding,
common-method diagnostics, multilevel models, and within-site sensitivity analyses
together yield a more credible picture than any single evidence source. Future
multi-site evaluations of educational technology should break the treatment--site
confound at the design stage by randomising within sites, and should pre-register both
their outcome families and their site-adjustment strategies.

The AutoSTEM project represents an initial step toward integrating SCCT and SDT in the
design and evaluation of secondary-school STEM interventions. By making PID control
accessible, engaging, and psychologically need-supportive for diverse learners,
interventions of this kind can contribute to building the inclusive, technically
competent workforce that the Fourth and Fifth Industrial Revolutions demand. The
gender gap finding, however, serves as a reminder that psychological need support
requires not only the {provision} of autonomy, competence, and relatedness, but
also vigilant attention to the {equitable distribution} of opportunities to
satisfy those needs across all participants.

\backmatter
\section*{Acknowledgments}
This work was supported by the Erasmus+ KA210-SCH programme under grant 2025-1-IT02-KA210-SCH-000358066 and Optimation AB. The authors want to thank the teachers and students from Luleå International School, Björkskataskolan, ITIS Mattei and Liceo Galilei for helping us with the workshops.

\bibliography{references_clean}

@inproceedings{morano2026serious,
  author    = {Morano, Martina and Collevecchio, Davide and Ripani, Alessandro and Giachini, Francesco and Gioiello, Flavia and Bartolucci, Veronica and Di Nardo, Francesco and Scaradozzi, David},
  title     = {A serious game approach for teaching control concepts from simulation-to-real robotic framework},
  booktitle = {2026 34th Mediterranean Conference on Control and Automation (MED)},
  year      = {2026},
  pages	= {745--750}
}

@article{Hughes2011CompetenceSelfEfficacy,
  title = {Perceived Competence: A Common Core for Self-Efficacy and Self-Concept?},
  volume = {93},
  ISSN = {1532-7752},
  url = {http://dx.doi.org/10.1080/00223891.2011.559390},
  DOI = {10.1080/00223891.2011.559390},
  number = {3},
  journal = {Journal of Personality Assessment},
  publisher = {Informa UK Limited},
  author = {Hughes,  Amanda and Galbraith,  David and White,  David},
  year = {2011},
  month = Apr,
  pages = {278–289}
}

@article{urrea2025industrial,
	title = {Industrial {Robotics} and {Adaptive} {Control} {Systems} in {STEM} {Education}: {Systematic} {Review} of {Technology} {Transfer} from {Industry} to {Classroom} and {Competency} {Development} {Framework}},
	volume = {16},
	doi = {10.3390/app16042026},
	number = {4},
	journal = {Applied Sciences},
	publisher = {Multidisciplinary Digital Publishing Institute},
	author = {Urrea, Claudio},
	year = {2026}
}

@article{chakraborty2024industry,
	title = {Industry 5.0 in {Smart} {Education}: {Concepts}, {Applications}, {Challenges}, {Opportunities}, and {Future} {Directions}},
	volume = {12},
	doi = {10.1109/ACCESS.2024.3401473},
	journal = {IEEE Access},
	author = {Supriya, Y. and Bhulakshmi, Dasari and Bhattacharya, Sweta and Gadekallu, Thippa Reddy and Vyas, Pratik and Kaluri, Rajesh and Sumathy, S. and Koppu, Srinivas and Brown, David J. and Mahmud, Mufti},
	year = {2024},
	pages = {81938--81967},
}

@article{lopez2023understanding,
	title = {Understanding {STEM} career choices: {A} systematic mapping},
	volume = {9},
	doi = {10.1016/j.heliyon.2023.e16676},
	number = {6},
	journal = {Heliyon},
	publisher = {Elsevier},
	author = {López, Pepa and Simó, Pep and Marco, Jordi},
	year = {2023},
}

@article{ribeirinha2024investigating,
	title = {Investigating the {Impact} of {STEM} {Inquiry}-{Based} {Learning} {Activities} on {Secondary} {School} {Student}’s {STEM} {Career} {Interests}: {A} {Gender}-{Based} {Analysis} {Using} the {Social} {Cognitive} {Career} {Framework}},
	volume = {14},
	doi = {10.3390/educsci14101037},
	number = {10},
	journal = {Education Sciences},
	publisher = {Multidisciplinary Digital Publishing Institute},
	author = {Ribeirinha, Teresa and Baptista, Mónica and Correia, Marisa},
	year = {2024},
	pages = {1037},
}

@article{kier2014development,
	title = {The {Development} of the {STEM} {Career} {Interest} {Survey} ({STEM}-{CIS})},
	volume = {44},
	doi = {10.1007/s11165-013-9389-3},
	number = {3},
	journal = {Research in Science Education},
	author = {Kier, Meredith W. and Blanchard, Margaret R. and Osborne, Jason W. and Albert, Jennifer L.},
	year = {2014},
	pages = {461--481},
}

@article{wang2023gender,
	title = {Gender differences in high school students’ interest in {STEM} careers: a multi-group comparison based on structural equation model},
	volume = {10},
	doi = {10.1186/s40594-023-00443-6},
	number = {1},
	journal = {International Journal of STEM Education},
	author = {Wang, Ning and Tan, Aik-Ling and Zhou, Xiaohong and Liu, Ke and Zeng, Feng and Xiang, Jiong},
	year = {2023},
	pages = {59},
}

@article{deci2008self,
	address = {US},
	title = {Self-determination theory: {A} macrotheory of human motivation, development, and health},
	volume = {49},
	doi = {10.1037/a0012801},
	number = {3},
	journal = {Canadian Psychology / Psychologie canadienne},
	publisher = {Educational Publishing Foundation},
	author = {Deci, Edward L. and Ryan, Richard M.},
	year = {2008},
	pages = {182--185},
}

@article{gilroy2026self,
	title = {Self-determination theory in the secondary school classroom: {A} scoping review},
	volume = {13},
	doi = {10.1016/j.ssaho.2026.102698},
	journal = {Social Sciences \& Humanities Open},
	author = {Gilroy, Aisling and McCloat, Amanda and Harvey, Deirdre},
	year = {2026},
	pages = {102698},
}

@article{ryan2000self,
	address = {US},
	title = {Self-determination theory and the facilitation of intrinsic motivation, social development, and well-being},
	volume = {55},
	doi = {10.1037/0003-066X.55.1.68},
	number = {1},
	journal = {American Psychologist},
	publisher = {American Psychological Association},
	author = {Ryan, Richard M. and Deci, Edward L.},
	year = {2000},
	pages = {68--78},
}

@article{ryan2020intrinsic,
	title = {Intrinsic and extrinsic motivation from a self-determination theory perspective: {Definitions}, theory, practices, and future directions},
	volume = {61},
	doi = {10.1016/j.cedpsych.2020.101860},
	journal = {Contemporary Educational Psychology},
	author = {Ryan, Richard M. and Deci, Edward L.},
	year = {2020},
	pages = {101860},
}

@article{vansteenkiste2020basic,
	title = {Basic psychological need theory: {Advancements}, critical themes, and future directions},
	volume = {44},
	doi = {10.1007/s11031-019-09818-1},
	number = {1},
	journal = {Motivation and Emotion},
	author = {Vansteenkiste, Maarten and Ryan, Richard M. and Soenens, Bart},
	year = {2020},
	pages = {1--31},
}

@article{ramirez2024evaluation,
	title = {Evaluation of the {Implementation} of {Project}-{Based}-{Learning} in {Engineering} {Programs}: {A} {Review} of the {Literature}},
	volume = {14},
	doi = {10.3390/educsci14101107},
	number = {10},
	journal = {Education Sciences},
	publisher = {Multidisciplinary Digital Publishing Institute},
	author = {Ramírez de Dampierre, Marta and Gaya-López, Maria Cruz and Lara-Bercial, Pedro J.},
	year = {2024},
	pages = {1107},
}

@article{sapi2025control,
	series = {14th {IFAC} {Symposium} on {Advances} in {Control} {Education} {ACE} 2025},
	title = {Control {Theory} with {Applications} in {Biomedical} {Engineering} for {Advanced} {STEM} {Students} in {High} {School}},
	volume = {59},
	doi = {10.1016/j.ifacol.2025.08.048},
	number = {7},
	journal = {IFAC-PapersOnLine},
	author = {Sápi, Johanna and Drexler, Dániel András and Eigner, György and Ferenci, Tamás and Kovács, Levente},
	year = {2025},
	pages = {207--212},
}

@article{dresser2024increasing,
	title = {Increasing {STEM} career interest: {The} role of out-of-school time {STEM} programs designed for underrepresented minorities},
	volume = {20},
	doi = {10.1371/journal.pone.0336418},
	number = {11},
	journal = {PLOS ONE},
	publisher = {Public Library of Science},
	author = {Dresser, Maria and Miller, Kelly and Sonnert, Gerhard and Sadler, Philip},
	year = {2025},
	pages = {e0336418},
}

@article{browning2025three,
	title = {A three-tiered mentorship approach for supporting high school students interested in science, technology, engineering, and mathematics ({STEM}) careers},
	volume = {9},
	doi = {10.1017/cts.2025.19},
	number = {1},
	journal = {Journal of Clinical and Translational Science},
	author = {Browning, Brittney D. and Glover, Janiece S. and Meredith, Lindsay R. and Kirkland, Anna E. and Gex, Kathryn S. and Tomko, Rachel L. and Duong, Felicity and Squeglia, Lindsay M.},
	year = {2025},
	pages = {e54},
}

@article{fabregues2023use,
	title = {Use of mixed methods research in intervention studies to increase young people’s interest in {STEM}: {A} systematic methodological review},
	volume = {13},
	doi = {10.3389/fpsyg.2022.956300},
	journal = {Frontiers in Psychology},
	publisher = {Frontiers},
	author = {Fàbregues, Sergi and Sáinz, Milagros and Romano, María José and Escalante-Barrios, Elsa Lucia and Younas, Ahtisham and López-Pérez, Beatriz-Soledad},
	year = {2023},
}

@article{parmaxi2023gender,
	title = {Gender equality in science, technology, engineering and mathematics: industrial vis-a-vis academic perspective},
	volume = {3},
	doi = {10.1007/s44217-023-00082-7},
	number = {1},
	journal = {Discover Education},
	author = {Parmaxi, Antigoni and Christou, Eirini and Fernández Valdés, Julia and Puente Hevia, Dalia María and Perifanou, Maria and Economides, Anastasios A. and Mazaj, Jelena and Manchenko, Maryna},
	year = {2024},
	pages = {3},
}

@article{stoet2018gender,
	title = {The {Gender}-{Equality} {Paradox} in {Science}, {Technology}, {Engineering}, and {Mathematics} {Education}},
	volume = {29},
	doi = {10.1177/0956797617741719},
	number = {4},
	journal = {Psychological Science},
	publisher = {SAGE Publications Inc},
	author = {Stoet, Gijsbert and Geary, David C.},
	year = {2018},
	pages = {581--593},
}

@article{ditullio2019gender,
	title = {Gender {Equality} in {STEM}: {Exploring} {Self}-{Efficacy} {Through} {Gender} {Awareness}},
	volume = {11},
	number = {Italian Journal of Sociology of Education 11/3},
	journal = {Italian Journal of Sociology of Education},
	publisher = {Padova University Press},
	author = {Tullio, Ilaria Di},
	year = {2019},
	pages = {226--245},
}

@article{hernandez2024challenge,
	title = {The {Challenge} of the {Gender} {Gap} and the {Lack} of {STEM} {Vocations} in {ESO} and {Baccalaureate}},
	volume = {19},
	doi = {10.1109/RITA.2024.3499337},
	journal = {IEEE Revista Iberoamericana de Tecnologias del Aprendizaje},
	author = {Hernández-Pérez, Mariana and Manuel Hernández-Castellano, Pedro and Alexis Alonso-Sánchez, José and Gregorio Quevedo-Gutiérrez, Eduardo},
	year = {2024},
	pages = {331--341},
}

@inproceedings{kovaleva2024pros,
	address = {New York, NY, USA},
	series = {{GE}@{ICSE} '22},
	title = {Pros and {Cons} of running educational hackathons in a gender-neutral fashion},
	isbn = {978-1-4503-9294-5},
	doi = {10.1145/3524501.3527603},
	booktitle = {Proceedings of the {Third} {Workshop} on {Gender} {Equality}, {Diversity}, and {Inclusion} in {Software} {Engineering}},
	publisher = {Association for Computing Machinery},
	author = {Kovaleva, Yekaterina and Happonen, Ari and Hasheela-Mufeti, Victoria},
	year = {2022},
	pages = {27--34},
}

@article{niemiec2009autonomy,
	title = {Autonomy, competence, and relatedness in the classroom: {Applying} self-determination theory to educational practice},
	volume = {7},
	doi = {10.1177/1477878509104318},
	number = {2},
	journal = {Theory and Research in Education},
	publisher = {SAGE Publications},
	author = {Niemiec, Christopher P. and Ryan, Richard M.},
	year = {2009},
	pages = {133--144},
}

@article{cheon2020autonomy,
	title = {When teachers learn how to provide classroom structure in an autonomy-supportive way: {Benefits} to teachers and their students},
	volume = {90},
	doi = {10.1016/j.tate.2019.103004},
	journal = {Teaching and Teacher Education},
	author = {Cheon, Sung Hyeon and Reeve, Johnmarshall and Vansteenkiste, Maarten},
	year = {2020},
	pages = {103004},
}

@article{howard2021student,
	title = {Student {Motivation} and {Associated} {Outcomes}: {A} {Meta}-{Analysis} {From} {Self}-{Determination} {Theory}},
	volume = {16},
	doi = {10.1177/1745691620966789},
	number = {6},
	journal = {Perspectives on Psychological Science},
	publisher = {SAGE Publications Inc},
	author = {Howard, Joshua L. and Bureau, Julien and Guay, Frédéric and Chong, Jane X. Y. and Ryan, Richard M.},
	year = {2021},
	pages = {1300--1323},
}

@article{leon2015self,
	title = {Self-determination and {STEM} education: {Effects} of autonomy, motivation, and self-regulated learning on high school math achievement},
	volume = {43},
	doi = {10.1016/j.lindif.2015.08.017},
	journal = {Learning and Individual Differences},
	author = {León, Jaime and Núñez, Juan L. and Liew, Jeffrey},
	year = {2015},
	pages = {156--163},
}

@article{andic2024direct,
	title = {Direct and indirect instruction in educational robotics: a comparative study of task performance per cognitive level and student perception},
	volume = {11},
	doi = {10.1186/s40561-024-00298-6},
	number = {1},
	journal = {Smart Learning Environments},
	author = {Anđić, Branko and Maričić, Mirjana and Mumcu, Filiz and Prodromou, Theodosia and Leoste, Janika and Saimon, Musa and Lavicza, Zsolt},
	year = {2024},
	pages = {12},
}

@article{nunez2025how,
	title = {How teacher autonomy support and student creativity jointly contribute to self-regulated learning: a dynamic, person-environment fit perspective},
	volume = {58},
	doi = {10.1016/j.tsc.2025.101873},
	journal = {Thinking Skills and Creativity},
	author = {Núñez-Regueiro, Fernando and Verger, Nicolas B. and Bressoux, Pascal and Batruch, Anatolia and Bouet, Marinette and Bressan, Marco and Brown, Genavee and Butera, Fabrizio and Cherbonnier, Anthony and Darnon, Céline and Demolliens, Marie and de Place, Anne-Laure and Desrichard, Olivier and Goron, Luc and Hémon, Brivael and Huguet, Pascal and Jamet, Eric and Mazenod, Vincent and Mella-Barraco, Nathalie and Michinov, Estelle and Michinov, Nicolas and Ofosu, Nana and Peter, Laurine and Poletti, Céline and Régner, Isabelle and Riant, Mathilde and Robert, Anaïs and Rudmann, Ocyna and Sanrey, Camille and Stanczak, Arnaud and Toumani, Farouk and Visintin, Emilio Paolo and Pansu, Pascal},
	year = {2025},
	pages = {101873},
}

@article{farrell_applying_2014,
	series = {Proceedings of the {European} {Conference} on {Games}-based {Learning}},
	title = {Applying the self determination theory of motivation in games based {Learning}: 8th {European} {Conference} on {Games} {Based} {Learning}},
	journal = {Proceedings of the 8th European Conference on Games Based Learning},
	publisher = {Academic Conferences and Publishing International Limited},
	author = {Farrell, David and Moffat, David},
	editor = {Busch, Carsten},
	year = {2014},
	pages = {118--127},
}

@article{suyatmo2025preparing,
	title = {Preparing industry-ready graduates: {Interplay} of lecturer behavior, self-motivation, and soft skill development in vocational education},
	volume = {261},
	doi = {10.1016/j.actpsy.2025.105892},
	journal = {Acta Psychologica},
	author = {Suyatmo, Suyatmo and Prayitno, Hadi and Wirawan, Willy Artha and Pratiwi, Damar Isti and Waluyo, Budi},
	year = {2025},
	pages = {105892},
}

@article{alston2026business,
	title = {Business-led informal {STEM} outreach: bridging industry, education, and workforce development},
	volume = {8},
	doi = {10.1186/s43031-026-00161-3},
	number = {1},
	journal = {Disciplinary and Interdisciplinary Science Education Research},
	author = {Alston-Socha, Willow and Rogers, Kristana and Hudson, Kesha and Campbell, Taylor},
	year = {2026},
	pages = {12},
}

@article{wang_autonomy-supportive_2025,
	title = {An autonomy-supportive intervention program for {STEM} teachers to enhance engagement among students},
	volume = {11},
	doi = {10.1016/j.heliyon.2025.e42150},
	number = {3},
	journal = {Heliyon},
	author = {Wang, Chee Keng John and Reeve, Johnmarshall and Liu, Woon Chia and Kee, Ying Hwa and Ng, Betsy and Chua, Li Lian and Kong, Leng Chee},
	year = {2025},
	pages = {e42150},
}

@article{patall2021daily,
  author  = {Patall, Erika A. and Zambrano, Jeanette and Kennedy, Alana A. U. and Yates, Nicole and Vall{\'\i}n, Jen A.},
  title   = {Daily, Weekly, and Cumulative Effects of Autonomy-Supportive Teaching in Secondary School Classrooms},
  journal = {Journal of Educational Psychology},
  volume  = {113},
  number  = {5},
  pages   = {1066--1087},
  year    = {2021},
  doi     = {10.1037/edu0000643},
}

@article{podsakoff2003common,
  author  = {Podsakoff, Philip M. and MacKenzie, Scott B. and Lee, Jeong-Yeon and Podsakoff, Nathan P.},
  title   = {Common Method Biases in Behavioral Research: A Critical Review of the Literature and Recommended Remedies},
  journal = {Journal of Applied Psychology},
  volume  = {88},
  number  = {5},
  pages   = {879--903},
  year    = {2003},
  doi     = {10.1037/0021-9010.88.5.879},
}

@book{raudenbush2002hierarchical,
  author    = {Raudenbush, Stephen W. and Bryk, Anthony S.},
  title     = {Hierarchical Linear Models: Applications and Data Analysis Methods},
  edition   = {2},
  publisher = {Sage Publications},
  address   = {Thousand Oaks, CA},
  year      = {2002},
}

@article{rosenbaum1983central,
  author  = {Rosenbaum, Paul R. and Rubin, Donald B.},
  title   = {The Central Role of the Propensity Score in Observational Studies for Causal Effects},
  journal = {Biometrika},
  volume  = {70},
  number  = {1},
  pages   = {41--55},
  year    = {1983},
  doi     = {10.1093/biomet/70.1.41},
}

@article{austin2011optimal,
  author  = {Austin, Peter C.},
  title   = {Optimal Caliper Widths for Propensity-Score Matching when Estimating Differences in Means and Differences in Proportions in Observational Studies},
  journal = {Pharmaceutical Statistics},
  volume  = {10},
  number  = {2},
  pages   = {150--161},
  year    = {2011},
  doi     = {10.1002/pst.433},
}

@book{shadish2002experimental,
  author    = {Shadish, William R. and Cook, Thomas D. and Campbell, Donald T.},
  title     = {Experimental and Quasi-Experimental Designs for Generalized Causal Inference},
  publisher = {Houghton Mifflin},
  address   = {Boston, MA},
  year      = {2002},
}

@article{duckworth2015measurement,
  author  = {Duckworth, Angela L. and Yeager, David S.},
  title   = {Measurement Matters: Assessing Personal Qualities Other than Cognitive Ability for Educational Purposes},
  journal = {Educational Researcher},
  volume  = {44},
  number  = {4},
  pages   = {237--251},
  year    = {2015},
  doi     = {10.3102/0013189X15584327},
}

@book{cohen1988statistical,
  author    = {Cohen, Jacob},
  title     = {Statistical Power Analysis for the Behavioral Sciences},
  edition   = {2},
  publisher = {Lawrence Erlbaum Associates},
  address   = {Hillsdale, NJ},
  year      = {1988},
}

@article{lakens2013calculating,
  author  = {Lakens, Dani{\"e}l},
  title   = {Calculating and Reporting Effect Sizes to Facilitate Cumulative Science: A Practical Primer for $t$-tests and ANOVAs},
  journal = {Frontiers in Psychology},
  volume  = {4},
  pages   = {863},
  year    = {2013},
  doi     = {10.3389/fpsyg.2013.00863},
}

\begin{appendices}

\section{Pre-Questionnaire}
\label{app:pre}

\setcounter{table}{0}
\setcounter{figure}{0}

\hfuzz=3pt
 
\definecolor{qgray}{gray}{0.93}
\definecolor{elsblue}{RGB}{0,84,166}
 
\newcommand{\cbox}{$\square$\;}

\newcolumntype{S}{>{\centering\arraybackslash}p{0.62cm}}

\newcommand{\qsec}[1]{\par\medskip\noindent{\bfseries #1}\par\smallskip\noindent}

\newcommand{\qrule}{\par\smallskip\noindent\textcolor{elsblue}{\rule{\linewidth}{0.5pt}}\par\smallskip}

\newcommand{\scalenote}[1]{%
  \par\smallskip
  \noindent\colorbox{qgray}{%
    \parbox{\dimexpr\linewidth-2\fboxsep\relax}{%
      \small\itshape #1}}%
  \par\medskip\noindent}

\newcommand{\likertheader}{%
  \rowcolor{qgray}%
  \textbf{Statement} & \textbf{1} & \textbf{2} & \textbf{3} & \textbf{4} & \textbf{5}\\}

\newcommand{\mcqq}[2]{%
  \par\smallskip\noindent\textbf{#1}\enspace #2\par\smallskip}

\newenvironment{mcqopts}{%
  \begin{enumerate}[label=\alph*),leftmargin=2em,itemsep=2pt,topsep=2pt,parsep=0pt]%
}{\end{enumerate}}

\qsec{A.1\quad Demographics}
\begin{tabular}{@{}ll@{\hspace{2.2cm}}ll@{}}
Age:         & \underline{\hspace{2.8cm}} &
Course year: & \underline{\hspace{2.8cm}} \\
\end{tabular}

\smallskip
\noindent\textbf{Gender:}\quad
\cbox Male\quad
\cbox Female\quad
\cbox Non-binary/other\quad
\cbox Prefer not to say

\medskip

\qsec{A.2\quad Background Experience}
Do you have experience in:

\smallskip
\begin{tabular}{@{}l@{\;}c@{\;}c@{\hspace{1.5cm}}l@{\;}c@{\;}c@{}}
Robotics          & \cbox Yes & \cbox No &
Automatic Control & \cbox Yes & \cbox No \\[5pt]
Programming       & \cbox Yes & \cbox No &
Serious game (simulator) & \cbox Yes & \cbox No \\
\end{tabular}

\medskip

\qsec{A.3\quad Video Game Usage}
How much time do you spend on average playing video games (consoles and PC)?

\smallskip
\begin{tabular}{@{}lcccc@{}}
\toprule
 & \small$<$1 h & \small 1--2 h & \small 2--4 h & \small$>$4 h \\
\midrule
\textbf{Weekdays} & \cbox & \cbox & \cbox & \cbox \\[4pt]
\textbf{Weekend}  & \cbox & \cbox & \cbox & \cbox \\
\bottomrule
\end{tabular}

\qrule
\scalenote{Questions A.4--A.6 use a 5-point Likert scale:\enspace
  \textbf{1}~=~Strongly disagree,\;
  \textbf{2}~=~Disagree,\;
  \textbf{3}~=~Neutral,\;
  \textbf{4}~=~Agree,\;
  \textbf{5}~=~Strongly agree.}

\qsec{A.4\quad Perceptions about Robotics and Control}
\begin{tabularx}{\linewidth}{@{}X S S S S S@{}}
\toprule\likertheader\midrule
I feel capable of understanding basic concepts of robotics & & & & & \\[4pt]
I feel capable of following an explanation about robotic systems even if the topic is new & & & & & \\[4pt]
I feel capable of applying robotics concepts in simple practical activities or simulations & & & & & \\[4pt]
I feel able to explain what a control system is & & & & & \\[4pt]
I feel able to explain how a PID controller works & & & & & \\[4pt]
I feel able to understand system behaviour by changing P, I, D & & & & & \\[4pt]
I believe that studying robotics can be useful for my future & & & & & \\
\bottomrule
\end{tabularx}

\medskip

\qsec{A.5\quad Interest}
\begin{tabularx}{\linewidth}{@{}X S S S S S@{}}
\toprule\likertheader\midrule
I am interested in topics related to automatic control & & & & & \\[4pt]
I am interested in robotics & & & & & \\[4pt]
I am interested in STEM (Science, Technology, Engineering, or Mathematics) subjects & & & & & \\[4pt]
I am interested in a career related to STEM fields & & & & & \\
\bottomrule
\end{tabularx}

\medskip

\qsec{A.6\quad Expectation}
\begin{tabularx}{\linewidth}{@{}X S S S S S@{}}
\toprule\likertheader\midrule
I think the activities will be engaging & & & & & \\[4pt]
I think the activity will help me better understand concepts related to automatic control & & & & & \\[4pt]
I think the activity will be challenging but manageable & & & & & \\[4pt]
I think the serious game (simulator) will be easy to use & & & & & \\[4pt]
I think the serious game (simulator) will help me better understand concepts of control and PID & & & & & \\[4pt]
I think learning with a serious game will be more interesting than a traditional lesson & & & & & \\
\bottomrule
\end{tabularx}

\qrule
\scalenote{Question A.7 is scored as:\enspace
  \textbf{1}~=~Correct,\;
  \textbf{0}~=~Not correct.}

\qsec{A.7\quad Knowledge}

\mcqq{Q1.}{A serious game (simulator) is:}
\begin{mcqopts}
  \item A game designed only for entertainment
  \item A game designed for educational or training purposes
  \item A video game with advanced graphics
  \item A game that requires physical activity
\end{mcqopts}

\mcqq{Q2.}{The main purpose of a control system is:}
\begin{mcqopts}
  \item To increase energy consumption
  \item To regulate a system to reach a desired output
  \item To make systems more complex
  \item To eliminate all disturbances completely
\end{mcqopts}

\mcqq{Q3.}{What does PID stand for?}
\begin{mcqopts}
  \item Proportional Integral Derivative
  \item Process Input Data
  \item Power Integrated Device
  \item Proportional Internal Difference
\end{mcqopts}

\mcqq{Q4.}{A PID controller uses:}
\begin{mcqopts}
  \item Only proportional action
  \item Only integral action
  \item Proportional, integral, and derivative actions
  \item Random adjustments based on system noise
\end{mcqopts}

\mcqq{Q5.}{What is meant by autonomous motion of a vehicle?}
\begin{mcqopts}
  \item Movement controlled manually by a human operator
  \item Movement that follows a predefined path without reacting to the environment
  \item Movement controlled by onboard systems using sensor feedback without human intervention
  \item Movement that is always random and unpredictable
\end{mcqopts}

\section{Post-Questionnaire}
\label{app:post}

\qsec{B.1\quad Demographics}
\begin{tabular}{@{}ll@{\hspace{2.2cm}}ll@{}}
Age:         & \underline{\hspace{2.8cm}} &
Course year: & \underline{\hspace{2.8cm}} \\
\end{tabular}

\smallskip
\noindent\textbf{Gender:}\quad
\cbox Male\quad
\cbox Female\quad
\cbox Non-binary/other\quad
\cbox Prefer not to say

\qrule
\scalenote{Questions B.2--B.4 use a 5-point Likert scale:\enspace
  \textbf{1}~=~Strongly disagree,\;
  \textbf{2}~=~Disagree,\;
  \textbf{3}~=~Neutral,\;
  \textbf{4}~=~Agree,\;
  \textbf{5}~=~Strongly agree.}

\qsec{B.2\quad Perceptions about Robotics and Control}
\begin{tabularx}{\linewidth}{@{}X S S S S S@{}}
\toprule\likertheader\midrule
I feel capable of understanding basic concepts of robotics & & & & & \\[4pt]
I feel capable of following an explanation about robotic systems even if the topic is new & & & & & \\[4pt]
I feel capable of applying robotics concepts in simple practical activities or simulations & & & & & \\[4pt]
I feel able to explain what a control system is & & & & & \\[4pt]
I feel able to explain how a PID controller works & & & & & \\[4pt]
I feel able to understand system behaviour by changing P, I, D & & & & & \\[4pt]
I believe that studying robotics can be useful for my future & & & & & \\
\bottomrule
\end{tabularx}

\medskip

\qsec{B.3\quad Interest}
\begin{tabularx}{\linewidth}{@{}X S S S S S@{}}
\toprule\likertheader\midrule
I am interested in topics related to automatic control & & & & & \\[4pt]
I am interested in robotics & & & & & \\[4pt]
I am interested in STEM (Science, Technology, Engineering, or Mathematics) subjects & & & & & \\[4pt]
I am interested in a career related to STEM fields & & & & & \\
\bottomrule
\end{tabularx}

\medskip

\qsec{B.4\quad Expectation}
\begin{tabularx}{\linewidth}{@{}X S S S S S@{}}
\toprule\likertheader\midrule
I think the activities were engaging & & & & & \\[4pt]
I think the activity helped me better understand concepts related to automatic control & & & & & \\[4pt]
I think the activity was challenging but manageable & & & & & \\[4pt]
I think the serious game (simulator) was easy to use & & & & & \\[4pt]
I think the serious game (simulator) helped me better understand concepts of control and PID & & & & & \\[4pt]
I think learning with a serious game was more interesting than a traditional lesson & & & & & \\
\bottomrule
\end{tabularx}

\qrule
\scalenote{Question B.5 is scored as:\enspace
  \textbf{1}~=~Correct,\;
  \textbf{0}~=~Not correct.}

\qsec{B.5\quad Knowledge}

\mcqq{Q1.}{A serious game (simulator) is:}
\begin{mcqopts}
  \item A game designed only for entertainment
  \item A game designed for educational or training purposes
  \item A video game with advanced graphics
  \item A game that requires physical activity
\end{mcqopts}

\mcqq{Q2.}{The main purpose of a control system is:}
\begin{mcqopts}
  \item To increase energy consumption
  \item To regulate a system to reach a desired output
  \item To make systems more complex
  \item To eliminate all disturbances completely
\end{mcqopts}

\mcqq{Q3.}{What does PID stand for?}
\begin{mcqopts}
  \item Proportional Integral Derivative
  \item Process Input Data
  \item Power Integrated Device
  \item Proportional Internal Difference
\end{mcqopts}

\mcqq{Q4.}{A PID controller uses:}
\begin{mcqopts}
  \item Only proportional action
  \item Only integral action
  \item Proportional, integral, and derivative actions
  \item Random adjustments based on system noise
\end{mcqopts}

\mcqq{Q5.}{What is meant by autonomous motion of a vehicle?}
\begin{mcqopts}
  \item Movement controlled manually by a human operator
  \item Movement that follows a predefined path without reacting to the environment
  \item Movement controlled by onboard systems using sensor feedback without human intervention
  \item Movement that is always random and unpredictable
\end{mcqopts}

\bigskip

\qsec{B.6\quad Open-Ended Questions}

\noindent\textbf{Q1.}\enspace What worked well during the activity?
  
\medskip
\noindent\textbf{Q2.}\enspace What did not work well? Suggestions?

\medskip
\noindent\textbf{Q3.}\enspace What is something interesting you learned?

\section{Open-Ended Response Analysis}
\label{app:open_ended}

The following tables present the thematic categorisation of responses to the three
open-ended questions on the post-questionnaire. Responses from all five Swedish
workshop sessions (Sessions~02--06, $N = 41$) are pooled. Each response was assigned
to a primary thematic category through manual coding.

\begin{table}[h!]
\centering
\caption{Thematic categorisation of open-ended responses to \textit{What worked well during the activity?} All five Swedish workshops pooled ($N = 41$).}
\label{tab:open_worked}
\begin{tabular}{lrr}
\toprule
\multicolumn{1}{c}{Theme} & \multicolumn{1}{c}{$n$} & \multicolumn{1}{c}{(\%)} \\
\midrule
    Practical / hands-on activities & 9 & (22\%) \\
    Collaboration / teamwork & 3 & (7\%) \\
    Clear instructions / materials & 3 & (7\%) \\
    Facilitator explanation quality & 2 & (5\%) \\
    Simulator / serious game & 4 & (10\%) \\
    Generic positive ("everything") & 11 & (27\%) \\
    Don't know / non-informative & 4 & (10\%) \\
    Nothing worked well & 3 & (7\%) \\
    Listening to presentation & 2 & (5\%) \\
\bottomrule
\end{tabular}
\end{table}

\begin{table}[h!]
\centering
\caption{Thematic categorisation of open-ended responses to \textit{What did not work well? Suggestions?} All five Swedish workshops pooled ($N = 41$). }
\label{tab:open_notworked}
\begin{tabular}{lrr}
\toprule
\multicolumn{1}{c}{Theme} & \multicolumn{1}{c}{$n$} & \multicolumn{1}{c}{(\%)} \\
\midrule
    Presentation too long / excessive theory & 5 & (12\%) \\
    Technical difficulties (connection, platform) & 8 & (20\%) \\
    Lack of conceptual clarity (PID, instructions) & 3 & (7\%) \\
    Too easy / insufficient challenge & 1 & (2\%) \\
    Too difficult / confusing & 3 & (7\%) \\
    Poor presentation quality / delivery & 3 & (7\%) \\
    Generic negative ("everything") & 3 & (7\%) \\
    Nothing / no issues & 8 & (20\%) \\
    Don't know / non-informative & 5 & (12\%) \\
    Wanted more activities & 1 & (2\%) \\
    Simulator/platform issues & 1 & (2\%) \\
\bottomrule
\end{tabular}
\end{table}

\begin{table}[h!]
\centering
\caption{Thematic categorisation of open-ended responses to \textit{What is something interesting you learned?} All five Swedish workshops pooled ($N = 41$).}
\label{tab:open_learned}
\begin{tabular}{lrr}
\toprule
\multicolumn{1}{c}{Theme} & \multicolumn{1}{c}{$n$} & \multicolumn{1}{c}{(\%)} \\
\midrule
    PID / control systems concepts & 7 & (17\%) \\
    Programming / how to program & 6 & (15\%) \\
    Robot behaviour / balance mechanisms & 4 & (10\%) \\
    Autonomy / autonomous systems & 1 & (2\%) \\
    Practical / engineering applications & 1 & (2\%) \\
    General STEM / robotics awareness & 3 & (7\%) \\
    Nothing / don't know & 14 & (34\%) \\
    Negative / explicit disinterest & 1 & (2\%) \\
    Meta-reflection on learning & 4 & (10\%) \\
\bottomrule
\end{tabular}
\end{table}

\section{Questionnaire to Dimension Mapping}
Table \ref{app:table_dimension} presents the mapping between questions in the pre- and post-questionnaires and the dimensions analyzed in this paper.

\label{app:table_dimension}
\begin{table}[h!]
\centering
\scriptsize
\setlength{\tabcolsep}{4pt}
\renewcommand{\arraystretch}{1.05}
\caption{Item-to-dimension mapping for SCCT constructs and SDT-aligned proxy dimensions.
Composite scores are unweighted means of constituent Likert items (1--5 scale).
Knowledge items are scored 0/1.}
\label{tab:item_mapping}
\begin{tabularx}{\textwidth}{@{}llX@{}}
\toprule
\textbf{Dimension} & \textbf{FW} & \textbf{Constituent Items} \\
\midrule
\multicolumn{3}{@{}l@{}}{\textit{SCCT dimensions}} \\[1pt]
Self-Efficacy        & SCCT & (SE1) Understanding basic robotics concepts; (SE2) following explanations about robotic systems; (SE3) applying robotics concepts in practical activities; (SE4) explaining what a control system is; (SE5) explaining how a PID controller works; (SE6) understanding system behaviour by varying P, I, D. \\[3pt]
Outcome Expectations & SCCT & (OE1) Studying robotics is useful for my future; (OE2) activity helped understand control concepts; (OE3) serious game helped understand control and PID. \\[3pt]
Interest             & SCCT & (IN1) Interest in automatic control; (IN2) interest in robotics; (IN3) interest in STEM subjects; (IN4) interest in a STEM-related career. \\[3pt]
Contextual Support   & SCCT & (CS1) Activities were engaging; (CS2) activity was challenging but manageable; (CS3) serious game was easy to use; (CS4) learning with serious game more interesting than traditional lesson. \\
\midrule
\multicolumn{3}{@{}l@{}}{\textit{SDT-aligned proxy dimensions}} \\[1pt]
Autonomy   & SDT  & CS3; CS2; CS1. \\[3pt]
Competence & SDT  & SE1--SE6. \\[3pt]
Relatedness          & SDT  & OE2; OE3; IN3; IN4. \\
\midrule
\multicolumn{3}{@{}l@{}}{\textit{Knowledge quiz (scored 0/1; not assigned to SCCT or SDT dimensions)}} \\[1pt]
Knowledge composite  & --   & (K1) Definition of a serious game; (K2) purpose of a control system; (K3) meaning of PID acronym; (K4) components of a PID controller; (K5) definition of autonomous motion. \\
\bottomrule
\end{tabularx}
\end{table}

\section{Additional Tables}
\begin{table}[h!]
\centering\small
\caption{Within-group pre--post changes (independent samples, Welch's $t$-test). Top panel: confirmatory outcomes (Holm-corrected within family). Bottom panel: secondary outcomes. $^{*}p<0.05$, $^{**}p<0.01$, $^{***}p<0.001$ (uncorrected).}
\label{tab:prepost}
\begin{tabular}{lrrrrrr}
\toprule
Outcome & SDT pre & Ctrl pre & & $\Delta$ / $g$ & & \\
\midrule
\multicolumn{7}{l}{\textbf{Primary (confirmatory) outcomes}} \\
\multicolumn{7}{l}{\textit{PID self-efficacy (composite)}} \\[1pt]
\quad $M_{pre}$ (SD) & 1.72 (0.79) & 1.53 (0.72) & $\Delta M_{SDT}$ & +0.98$^{***}$ & $\Delta M_{Ctrl}$ & +1.16$^{***}$ \\
\quad Hedges' $g$ & \multicolumn{2}{c}{+1.01} & \multicolumn{2}{c}{+1.25} & $p_{Holm}$ (SDT) & $<0.001$ \\[3pt]
\multicolumn{7}{l}{\textit{Knowledge quiz (0--5)}} \\[1pt]
\quad $M_{pre}$ (SD) & 3.31 (1.40) & 2.32 (1.49) & $\Delta M_{SDT}$ & +0.38 & $\Delta M_{Ctrl}$ & +0.26 \\
\quad Hedges' $g$ & \multicolumn{2}{c}{+0.27} & \multicolumn{2}{c}{+0.20} & $p_{Holm}$ (SDT) & 0.358 \\[3pt]
\multicolumn{7}{l}{\textit{STEM interest (composite)}} \\[1pt]
\quad $M_{pre}$ (SD) & 2.65 (1.05) & 3.28 (0.61) & $\Delta M_{SDT}$ & -0.09 & $\Delta M_{Ctrl}$ & +0.01 \\
\quad Hedges' $g$ & \multicolumn{2}{c}{-0.09} & \multicolumn{2}{c}{+0.02} & $p_{Holm}$ (SDT) & 0.669 \\[3pt]
\midrule
\multicolumn{7}{l}{\textbf{Secondary outcomes}} \\
\multicolumn{7}{l}{\textit{SCCT self-efficacy (composite)}} \\[1pt]
\quad $M_{pre}$ (SD) & 2.71 (0.66) & 2.35 (0.58) & $\Delta M_{SDT}$ & +0.55$^{**}$ & $\Delta M_{Ctrl}$ & +0.53$^{*}$ \\
\quad Hedges' $g$ & \multicolumn{2}{c}{+0.70} & \multicolumn{2}{c}{+0.72} & $p_{Holm}$ (SDT) & 0.008 \\[3pt]
\multicolumn{7}{l}{\textit{Robotics perception (composite)}} \\[1pt]
\quad $M_{pre}$ (SD) & 2.80 (0.67) & 2.51 (0.53) & $\Delta M_{SDT}$ & +0.45$^{**}$ & $\Delta M_{Ctrl}$ & +0.50$^{*}$ \\
\quad Hedges' $g$ & \multicolumn{2}{c}{+0.59} & \multicolumn{2}{c}{+0.74} & $p_{Holm}$ (SDT) & 0.029 \\[3pt]
\multicolumn{7}{l}{\textit{SCCT outcome expectations}} \\[1pt]
\quad $M_{pre}$ (SD) & 3.46 (0.92) & 3.32 (0.50) & $\Delta M_{SDT}$ & -0.10 & $\Delta M_{Ctrl}$ & +0.30 \\
\quad Hedges' $g$ & \multicolumn{2}{c}{-0.11} & \multicolumn{2}{c}{+0.51} & $p_{Holm}$ (SDT) & 1.000 \\[3pt]
\multicolumn{7}{l}{\textit{SCCT contextual support}} \\[1pt]
\quad $M_{pre}$ (SD) & 3.44 (0.78) & 3.24 (0.57) & $\Delta M_{SDT}$ & -0.09 & $\Delta M_{Ctrl}$ & +0.33 \\
\quad Hedges' $g$ & \multicolumn{2}{c}{-0.11} & \multicolumn{2}{c}{+0.52} & $p_{Holm}$ (SDT) & 1.000 \\[3pt]
\multicolumn{7}{l}{\textit{Perceived engagement}} \\[1pt]
\quad $M_{pre}$ (SD) & 3.69 (0.94) & 3.37 (0.68) & $\Delta M_{SDT}$ & -0.35 & $\Delta M_{Ctrl}$ & +0.89$^{***}$ \\
\quad Hedges' $g$ & \multicolumn{2}{c}{-0.35} & \multicolumn{2}{c}{+1.31} & $p_{Holm}$ (SDT) & 0.308 \\[3pt]
\bottomrule
\end{tabular}
\end{table}

\end{appendices}



\end{document}